\documentclass[10pt]{article}
\usepackage{amsmath}
\usepackage{latexsym}
\usepackage{amssymb}
\usepackage{amsfonts}
\usepackage{amsthm}
\title{CONTROL THEORY AND PARAMETRIZATIONS OF \\ 
LINEAR PARTIAL DIFFERENTIAL OPERATORS}
\author{J.-F. POMMARET \\ jean-francois.pommaret@wanadoo.fr \\
ORCID: 0000-003-0907-2601 }
\date{  }
\begin{document}
\maketitle

\noindent
{\bf ABSTRACT}    \\

\noindent
When ${\cal{D}}:\xi \rightarrow \eta$ is a linear ordinary differential (OD) or partial differential (PD) operator, a "direct problem" is to find the generating compatibility conditions (CC) in the form of an operator ${\cal{D}}_1:\eta \rightarrow \zeta$ such that ${\cal{D}}\xi=\eta$ implies ${\cal{D}}_1\eta=0$. When ${\cal{D}}$ is involutive, the procedure provides successive first order involutive operators ${\cal{D}}_1, ... , {\cal{D}}_n$ when the ground manifold has dimension $n$. Conversely, when ${\cal{D}}_1$ is given, a much more difficult " inverse problem " is to look for an operator ${\cal{D}}: \xi \rightarrow \eta$  having the generating CC ${\cal{D}}_1\eta=0$. If this is possible, that is when the differential module defined by 
${\cal{D}}_1$ is " {\it torsion-free} ", that is when it does not exist any observable quantity which is a sum of derivatives of ${\eta}$ that could be a solution of an {\it autonomous} OD or PD equation {\it for itself}, one shall say that the operator ${\cal{D}}_1$ is parametrized by ${\cal{D}}$. The parametrization is said to be "{\it minimum} "  if the differential module defined by ${\cal{D}}$ does not contain any free differential submodule. The systematic use of the adjoint of a differential operator provides a constructive test  with five steps.  We prove and illustrate through many explicit examples the fact that a control system  is controllable {\it if and only if} it can be parametrized. Accordingly, the controllability of any OD or PD control system is a " {\it built in} " property not depending on the choice of the input and output variables among the system variables. In the OD case and when ${\cal{D}}_1$ is formally surjective, controllability just amounts to the injectivity  of $ad({\cal{D}}_1)$, even in the variable coefficients case, a result still not  acknowledged by the control community. Among other applications, the parametrization of the Cauchy stress operator in arbitrary dimension $n$ has attracted many famous scientists (G.B. Airy in 1863 for $n=2$, J.C. Maxwell in 1863, G. Morera and E. Beltrami in 1892 for $n=3$, A. Einstein in 1915 for $n=4$). We prove that all these works are already explicitly using the self-adjoint Einstein operator, {\it which cannot be parametrized}. As a byproduct, they are all based on a confusion between the so-called $div$ operator induced from the Bianchi operator ${\cal{D}}_2$ and the Cauchy operator, adjoint of the Killing operator ${\cal{D}}$ which is parametrizing the Riemann operator ${\cal{D}}_1$ for an arbitrary $n$. We prove that this purely mathematical result deeply questions the origin and existence of gravitational waves. Like the Michelson and Morley experiment, it is thus an open historical problem to know whether Einstein was aware of these  previous works or not, as the comparison needs no comment.   \\

 \vspace{4cm}


\newpage

\noindent
1) {\bf INTRODUCTION}:\\

As will be shown in the many explicit examples presented in this paper, the solution space of many systems of ordinary differential (OD) or partial differential (PD) equations in engineering or mathematical physics "{\it can or cannot}" be parametrized by a certain number of arbitrary functions behaving like "{\it potentials}". More precisely, if a linear inhomogeneous system of OD or PD equations is given in the form ${\cal{D}}\xi=\eta$ where $\cal{D}$ is a differential operator acting on a certain number of functions $\xi$ in such a way as to provide a certain number of functions $\eta$ as second members, a {\it direct problem} is to know about the (generating) {\it compatibility conditions} (CC) in the form ${\cal{D}}_1\eta=0$ that {\it must} be satisfied by $\eta$ in general and the solution of such a problem has been known since a long time [1, 2]. Conversely, when a homogeneous system is given in the form ${\cal{D}}_1\eta=0$, the {\it inverse problem} is to decide whether {\it there exists or not} an operator $\cal{D}$ such that, if we write formally ${\cal{D}}\xi=\eta$, then the CC for $\eta$ are generated by ${\cal{D}}_1\eta=0$ and the solution of such a problem has only been discovered recently [3, 4, 5, 6, 7, 8, 9]. Both the direct and inverse problems can now be solved by computer algebra using Janet, Gr\"{o}bner or Pommaret bases.\\

In view of the examples to be met later on, it is important to notice that the parametrizing operator  may be of high order. Among the well known examples, we recall that a classical OD {\it control system} is {\it parametrizable} if and only if it is {\it controllable} (Kalman test of 1969 in [10]). Among PD systems, the electromagnetic (EM) {\it field}, solution of the first set of four Maxwell equations, admits a well known first order parametrization by means of the EM {\it potential} while the EM {\it induction}, solution of the second set of four Maxwell equations, also admits a first order parametrization by means of the so-called EM {\it pseudopotential}. On the contrary, it is now known since $1995$ [3, 10] that the set of ten second order linearized Einstein equations for the ten perturbtion of the metric cannot be parametrized and {\it cannot} therefore be considered as field equations (see [9] for more details and http://wwwb.math.rwth-aachen.de/OreModules for a computer  algebra solution). One among the best interesting and useful cases is concerned with continuum mechanics where the first order stress equations (in vacuum) admits a rather simple second order parametrization by means of the single Airy function in dimension 2 and, as we shall see later on, a much more complicated second order parametrization can be achieved in dimension $n\geq 2$.\\

It is also now known that all the above problems are particular cases of a {\it more sophisticated} and general situation involving the {\it formal theory} of systems of PD equations pioneered by D.C. Spencer an collaborators after 1960 [11, 12] (jet theory, diagram chasing, differential sequences,...)(see [1, 2] for more details) and {\it differential modules} in the framework of "{\it algebraic analysis}" pioneered in $1970$ by V. Palamodov [13] and, after $1990$, by U. Oberst [14] for the constant coefficients case and by M. Kashiwara [4]  for the variable coefficients case, without ever providing any explicit example (See [6] for more details or examples and also consult Zentralblatt Zbl 1079.93001 for a review). The corresponding {\it differential duality theory,} that is at the heart of all the previous examples and will be a central tool in this paper, highly depends on {\it homological algebra techniques} (localization, resolutions, extension modules,...) {\it which cannot be avoided}. \\

The purpose of the next sections is to apply these techniques in a way as simple and self-contained as possible in order to give a positive and explicit answer concerning the possibility to exhibit a {\it first order} parametrization of the stress/couple-stress equations met in the study of  Cosserat media. At the same time, as a corollary of the homological test, we shall give {\it for the first time} the reason for which the CC for the deformation tensor in classical elasticity theory are {\it second order} while the corresponding CC for Cosserat fields [15] are only {\it first order} and explain why this order is equal to the order of the corresponding parametrization. \\

At the end of the paper, we shall give hints in order to explain why, though the "{\it fields}" and their CC in classical and Cosserat elasticity theories look like {\it completely different at first sight}, therefore providing different presentations of the corresponding field equations, nevertheless {\it the possibility to obtain a parametrization in one framework necessarily implies the possibility to have a parametrization in the other framework and vice-versa}, {\it even in the variable coefficients case}. Though striking it may look like in such an engineering background, this {\it totally not evident result}, which is either known nor acknowledged today), is one of the simplest consequences of a basic result of homological algebra. In particular {\it the reader must look at the next section below with care, even though it does not seem to have anything to do with Maxwell or Cosserat equations}. At the same time, revisiting the work of H. Weyl on electromagnetism in the light of group theory, exactly as we did for the work of E. and F. Cosserat on elasticity [16, 17, 18], we shall point out the close relation existing between the second set of Maxwell equations and the Cosserat equations, both with their parametrizations. Our claim is that one can treat OD or PD examples in a unique framework, {\it on the condition to revisit almost entirely the classical OD case} because we shall understand why {\it  the controllability of a control system is a " built in " property not dependent on the presentation of the system or even on the choice of the input and output variables among the system variables}, a result quite far from what is believed today by any control engineer. To be more convincing, we ask the reader to realize the double pendulum experiment with a few dollars and to try to imagine what can be the link with the Cosserat and Einstein equations !.  \\

\noindent
2)  {\bf MECHANICAL MOTIVATIONS}   \\

     In the middle of the last century, {\it commutative algebra}, namely the study of modules over rings, was facing a very subtle problem, the resolution of which led to the modern but difficult {\it homological algebra}. Roughly, the problem was essentially to study properties of finitely generated modules not depending on the "{\it presentation}" of these modules by means of generators and relations. This "{\it hard step}" is based on homological/cohomological methods like the so-called "{\it extension}" modules which cannot therefore be avoided ([6, 19, 20] are fine references).\\

A classical OD control system {\it must} be brought to first order with no derivative of input in order to apply the well known Kalman test for checking its controllability or, equivalently, the possibility to parametrize it. However, there may be many different ways for following such a procedure and not a word is left for systems of PD equations. It is only after 1990 that a general OD/PD test has been provided, showing that {\it controllability is a "built in" property of a control system} as we already said, contrary to engineering intuition [4, 6, 7, 10, 4].\\
     
 As before, using now rings of "{\it differential operators}" instead of polynomial rings leads to {\it differential modules} and to the challenge of adding the word "{\it differential}" in front of concepts of commutative algebra. Accordingly, not only one needs properties not depending on the presentation, as we just explained, but also properties not depending on the coordinate system as it becomes clear from any application to mathematical or engineering physics where tensors and exterior forms are always to be met like in the space-time formulation of electromagnetism or General Relativity. Unhappily, no one of the previous techniques for OD/PD equations could work !. \\
     
By chance, the intrinsic study of systems of OD/PD equations has been pioneered in a totally independent way by D. C. Spencer and collaborators as we said, in a way superseding the "{\it leading term}" approach of Janet in 1920 [21] or Gr\"{o}bner in 1940 but quite poorly known by the mathematical community, even today. Accordingly, it was another challenge to unify the "{\it purely differential}" approach of Spencer with the "{\it purely algebraic}" approach of commutative algebra, having in mind the necessity to use the previous homological algebraic results in this new framework. This sophisticated mixture of differential geometry and homological algebra, now called "{\it algebraic analysis}" or " {\it differential homological algebra  ", } has been achieved between $1970$ and $1990$ as we shall explain.\\

Let $k$ be a field containing the subfield $\mathbb{Q}$ of rational numbers and $\chi=({\chi}_1,...,{\chi}_n)$ be indeterminates over $k$. We denote by $A=k[{\chi}_1,...,{\chi}_n]$ the ring of polynomials with coefficients in $k$. Next, let us introduce $n$ commuting derivations $d_1,...,d_n$ for which $k$ should be a field of constants and define the ring $D=k[d]=k[d_1,...,d_n]$ of differential operators with coefficients in $k$. Then $D$ and $A$ are isomorphic by $d_i\leftrightarrow {\chi}_i$. However, the (non-commutative) situation for a differential field $K$ with $n$ commuting derivations ${\partial}_1,...,{\partial}_n$ and subfield of constants $k$ escapes from the previous (commutative) approach and must be treated "{\it by its own}". For this, let $\mu=({\mu}_1,...,{\mu}_n)$ be a multi-index with {\it length} $\mid\mu\mid={\mu}_1+ ... +{\mu}_n$. We set $\mu +1_i=({\mu}_1,...,{\mu}_{i-1},{\mu}_i+1,{\mu}_{i+1},...,{\mu}_n)$ and we say that $\mu$ is of {\it class} $i$ if ${\mu}_1=...={\mu}_{i-1}=0, {\mu}_i\neq 0$. Accordingly, any operator $P=a^{\mu}d_{\mu}\in D$ acts on the (formal) unknowns $y^k$ for $k=1,...,m$ as we may set $d_{\mu}y^k=y^k_{\mu}$ with $y^k_0=y^k$ and introduce formally the {\it jet coordinates} $y_q=\{y^k_{\mu}\mid k=1,...,m; 0\leq \mid \mu \mid \leq q\}$. A system of PD equations can be written in the form ${\Phi}^{\tau}\equiv a^{\tau\mu}_ky^k_{\mu}=0$ with $a\in K $ and we define the (formal) {\it prolongation} of ${\Phi}^{\tau}$ with respect to $d_i$ to be $d_i{\Phi}^{\tau}\equiv a^{\tau\mu}_ky^k_{\mu +1_i}+ {\partial}_ia^{\tau\mu}_ky^k_{\mu}$. Finally, setting $Dy=Dy^1+...+Dy^m\simeq D^m$, we may introduce the differential module $M=Dy/D\Phi$ and induce maps $d_i:M\rightarrow M: {y}^k_{\mu}\rightarrow {y}^k_{\mu +1_i}$ by residue.\\

First of all, setting as usual $d=d_1=d/dt=dot $ when $n=1$, we sketch the technique of "{\it localization}" in the case of OD equations, comparing to the situation met in classical control theory. If we have a given system of OD equations, a basic question in control theory is to decide whether the control system is "{\it controllable}" or not. It is not our purpose to discuss here about such a question from an engineering point of view, but we just want to provide the algebraic counterpart in terms of a property of the corresponding differential module. We explain our goal on an academic example.\\

\noindent
{\bf EXAMPLE 2.1}: With $m=3$ and a constant parameter $a$, we consider the first order system ${\Phi}^1\equiv {\dot{y}}^1-ay^2-{\dot{y}}^3=0, {\Phi}^2\equiv y^1-{\dot{y}}^2+{\dot{y}}^3=0$. In order to study the {\it transfer matrix}, the idea is to replace the Laplace transform by another purely formal technique that could also be useful for studying systems of PD equations with variable coefficiets. For this, let us replace "{\it formally}" $d$ by the purely algebraic symbol $\chi$ whenever it appears and obtain the system of {\it linear equations} :\\
\[ \chi y^1-ay^2-\chi y^3=0, 1y^1-\chi y^2+\chi y^3=0 \Rightarrow y^1=\frac{\chi (\chi +a)}{{\chi}^2-a} y^3, y^2=\frac{\chi (\chi +1)}{{\chi}^2-a} y^3 \]
but we could have adopted a different choice for the only arbitrary unknown used as single input. At this step there are only two possibilities :\\
$\bullet \,\, a\neq 0,1 \Rightarrow $no "{\it simplification}" may occur and, getting rid of the common denominator, we get an algebraic parametrization leading to a differential parametrization as follows:\\
\[ y^1=\chi (\chi +a) z, y^2=\chi (\chi +1) z, y^3=({\chi}^2-a)z \Rightarrow y^1=\ddot{z}+a\dot{z}, y^2=\ddot{z}+\dot{z}, y^3=\ddot{z}-az  \]
$\bullet \,\, a=0$ or $a=1 \Rightarrow$  a "{\it simplification}" may occur and no parametrization can be found. For example, if $a=0$, setting $z=y^1-y^3$ we get $\chi z=0$ that is to say 
$\dot{z}=0$, while, if $a=1$, setting $z'=y^1 - y^2$, we get $(\chi - 1)z'=0$ that is to say $dz' - z'=0$.   \\
Though similar examples could be found in any textbook on control theory, it does not seem that such a procedure could bring any distinction between the two conditions $a=0$ and $a=1$. It is only quite later on that we shall understand the difference existing between these two conditions by revisiting the present example.   \\

\noindent
{\bf EXAMPLE 2.2}:  With $n=1, K= \mathbb{Q} $, let us consider the differential module $N$ defined by the OD equation $ \Phi \equiv (d^5 v + 3 d^4 v - 4 d^2 v) - (d^3 u - d u)$, that is $N= (Du + Dv)/D\Phi$. We may define the input differential module $L =Du$ by using $u$ and the output differential module $M=Dy \subset Dv$ by setting $y=d^2 v$. The differential module $(L + M) \subset M$ with a strict inclusion, is defined by the OD equation $\Psi \equiv (d^3 y + 3 d^2 y - 4 y) -(d^3 u - du) = 0$ that we can also write 
$(d-1)((d+2)^2 y - d (d+1) u) = 0$ because $K$ is a field of constants. As we can factor by $(d-1)$ it follows that $t(L + M)$ is generated by $z= (d^2 y + 4 dy + 4y) - (d^2 u + du)$ that satisfies $dz - z=0$. The annihilator ideal of $N/L$ is $ann(N/L)=(d^2 (d -1)(d+2)^2)$ and its radical is $(d(d -1)(d+2))=(d) \cap (d-1) \cap (d+2)$ which is an intersection of prime ideals. Similarly, we have $y=0 \Rightarrow d^2v=0, (d^3 -d)u=0$ and thus $ann(N/M)= (d^2) \cap (d^3 - d)$ leading to $rad(ann(N/M))= (d) \cap (d-1) \cap (d+1)$ which is also an intersection of prime differential ideals.   \\

\noindent
{\bf REMARK 2.3}: We have proved in ([6, 10]) how to use these differential submodules of $N$ both with the new differential modules $L'=L + t(N)$ and $M'= M + t(N) $ in order to study 
{\it all} the problems concerning poles and zeros of control systems. As we are only interested by controllability, we have just to study the differential submodules of the torsion-free differential module $N/t(N)$. If we suppose that $L\cap M =0$, we have the following commutative diagram of inclusions, in which the upper commutative square is the so-called 
{\it minimum controllable realization}:

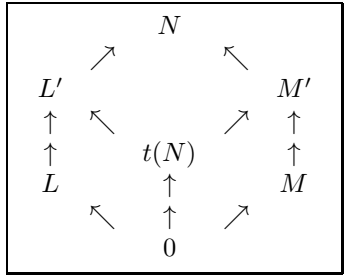
\[ \fbox{  $  \begin{array}{ccccc}
  &  &  N  &  &  \\
 & \nearrow &  & \nwarrow  &    \\
L' &  &  &  &  M'     \\
\uparrow  &  \nwarrow  &  & \nearrow &  \uparrow  \\
\uparrow  &   &  t(N)  &  &  \uparrow  \\  
  L  &   & \uparrow   &   &  M    \\
   &   \nwarrow  & \uparrow &  \nearrow  &  \\
   &  &  0  &  &  
\end{array}   $  }   \]
and just need to use the following delicate proposition [6]:   \\

\noindent
{\bf PROPOSITION 2.4}: If $0 \rightarrow M' \rightarrow M \rightarrow M'' \rightarrow 0$ is the short exact sequence of (differential) modules, then we have the formula 
$ rad(ann(M)) = rad(ann(M')) \cap rad(ann(M''))$ in which the radical of a (differential) ideal $\mathfrak{a}$ is the (differential) ideal generated by all the elements with a power in 
$\mathfrak{a}$.  \\

Recapitulating, we discover that a control system is controllable and thus parametrizable if and only if one cannot get any {\it autonomous element} satisfying an OD equation {\it by itself}. For understanding such a result in an algebraic manner, let $M$ be a module over an integral domain $A$ containing 1. A subset $S\subset A$ is called a {\it multiplicative subset} if $1\in S$ and $\forall s,t \in S \Rightarrow st \in S$. Moreover, we shall need/use the {\it Ore condition} on $S$ and $A$, namely $aS\cap sA\neq \emptyset , \forall a\in A, s\in S$.\\

\noindent
{\bf DEFINITION 2.5}: The {\it localization} of $M$ at $S$ is $S^{-1}M=\{s^{-1}x{\mid}s\in S,x\in M/\sim\}$ with $s^{-1}x\sim t^{-1}y \Leftrightarrow \exists u,v\in A, us=vt\in S , ux=vy$ (reduction to the same denominator in $S$) and we may introduce $t_S(M)=\{x\in M\mid \exists s\in S, sx=0\} $ as the kernel of the morphism $ M \rightarrow S^{-1}M : x \rightarrow 1^{-1}x$. If $S=A-\{ 0 \} \Rightarrow S^{-1}A=Q(A) $ field of fractions of $A$ and we introduce the {\it torsion submodule} $t_S(M)=t(M)=\{ x\in M \mid \exists 0\neq a\in A, ax=0\}$ of $M$. \\

In the case of a torsion-free module, that is when $t(M)=0$, reducing to the same denominator as in the control example or as in the next example, we have the following classical proposition amounting to exhibit a parametrization. However, the reader must notice that {\it it is useless in actual practice} as one needs a test (like the Kalman test) for checking the torsion-free condition. {\it This will be the hard part of the job in this paper} !.\\

\noindent
{\bf PROPOSITION 2.6}: When $M$ is a finitely generated torsion-free module and $S=A-\{0\}$, from the inclusion of $M$ into the vector space $S^{-1}M$ over $Q(A)$, we deduce that there exists a finitely generated free module $F$ over $A$ with $M\subset F$.\\ 

\noindent
{\bf EXAMPLE 2.7}: As an unexpected application to $2$-dimensional elasticity, let us consider the well known Cauchy stress equations:     \\
\[  \fbox{  $  {\partial}_1 {\sigma}^{11}+{\partial}_2{\sigma}^{21}=0, \, \, \,  {\partial}_1{\sigma}^{12}+{\partial}_2{\sigma}^{22}=0  $  }  \]
with ${\sigma}^{12}={\sigma}^{21}$. Replacing ${\partial}_i$ by ${\chi}_i$, we may localize and obtain:\\
\[    {\chi}_1{\sigma}^{11}+{\chi}_2{\sigma}^{21}=0, {\chi}_1{\sigma}^{12}+{\chi}_2{\sigma}^{22}=0  \]
Reducing the fractions to the same denominator, we get:\\
\[  {\sigma}^{11}=-\frac{{\chi}_2}{{\chi}_1}{\sigma}^{21}=-\frac{({\chi}_2)^2}{{\chi}_1{\chi}_2}{\sigma}^{12}, {\sigma}^{22}=-\frac{{\chi}_1}{{\chi}_2}{\sigma}^{12}=-\frac{({\chi}_1)^2}{{\chi}_1{\chi}_2} {\sigma}^{12} \]
and obtain therefore the $1$-dimensional subvector space over $\mathbb{Q}({\chi}_1,{\chi}_2)$:\\
\[ {\sigma}^{11}=({\chi}_2)^2\phi, \,\, {\sigma}^{12}={\sigma}^{21}=-{\chi}_1{\chi}_2\phi, \,\, {\sigma}^{22}=({\chi}_1)^2\phi \]
a result providing at once the well known parametrization by the Airy function:\\
\[   \fbox{  $ {\sigma}^{11}={\partial}_{22}\phi, \,\, {\sigma}^{12}={\sigma}^{21}=-{\partial}_{12}\phi, \,\, {\sigma}_{22}={\partial}_{11}\phi  $  }  \]
It may be interesting to compare this purely formal approach to the standard analytic aproach presented in any textbook along the following way. From the first stress equation and Stokes identity for the curl, there exists a function $\varphi$ such that ${\sigma}^{11}={\partial}_2\varphi, {\sigma}^{21}=-{\partial}_1\varphi$. Similarly, from the second stress equation, there exists a function $\psi$ such that ${\sigma}^{22}={\partial}_1\psi, {\sigma}^{12}= - {\partial}_2\psi$. Finally, from the symmetry of the stress, there exists a function $\phi$ such that $\varphi={\partial}_2\phi, \psi={\partial}_1\phi$ and we find back the same parametrization of course. The reader must notice that, in this example, one can check that the parametrization does work but no geometric inside can be achieved in arbitrary dimension $n\geq 2$, even though exactly the same procedure can be applied through computer algebra (see A. Quadrat in http://www.risc.uni-linz.ac.at/about/conferences/aaca09/ModuleTheoryI.pdf).\\

Taking into account the works of Janet [21] and Spencer [11], the study of systems of PD equations cannot be achieved without understanding {\it involution} and we now explain this concept by exhibiting the usefu " {\it Janet tabular} ". For this, changing linearly the derivations if necessary, we may successively solve the maximum number of equations with respect to the jets of order $q$ and class $n$, class $(n-1)$,..., class 1. Moreover, for each equation of order $q$ and class i, $d_1,...,d_i$ are called {\it multiplicative} while $d_{i+1},...,d_n$ are called {\it nonmultiplicative} and $d_1,...,d_n$ are nonmultiplicative for all the remaining equations of order $\leq q-1$.\\

\noindent
{\bf DEFINITION 2.8}: The system is said to be {\it involutive} if each prolongation with respect to a nonmultiplicative derivation is a linear combination of prolongations with respect to the multiplicative ones  [1, 2, 6].\\

\noindent
{\bf EXAMPLE 2.9}: The system $y_{11}=0, y_{13}-y_2=0$ is {\it not} involutive. Effecting the permutation $(1,2,3)\rightarrow (3,2,1)$, we get the system $y_{33}=0, y_{13}-y_2=0$. As $d_1y_{33}-d_3(y_{13}-y_2)=y_{23}$ and $d_1y_{23}-d_2(y_{13}-y_2)=y_{22}$, the system $y_{33}=0, y_{23}=0, y_{22}=0, y_{13}-y_2=0$ is involutive with 1 equation of class 3, 2 equations of class 2 and 1 equation of class 1. Another tricky example provided for $n=3$ by Macaulay in  [22] is $y_{33}=0, y_{23} - y_{11}=0, y_{22}=0)$ with $8 = 2^3$ parametric jet coordinates $(y, y_1, y_2, y_3, y_{11}, y_{12}, y_{13}, y_{111})$.  \\

\noindent
{\bf EXAMPLE 2.10}: The Killing system $y^j_i+y^i_j=0$ for the Eucldean metric is {\it not} involutive but the first prolongation $y^j_i+y^i_j=0, y^k_{ij}=0$ is involutive. This is the reason for which the Riemann tensor is a first order expression in the metric and Christoffel symbols and thus second order in the metric alone (for more details, see [2], p 249-258).\\

\noindent
{\bf APPLICATION 2.11}: $t(M)=M$ if and only if the number of equations of class $n$ is $m$. Otherwise there is a strict inclusion $t(M)\subset M$ and, when $t(M)=0$, the minimum number of potentials in any parametrization is equal to the number of unknowns minus the number of equations of class $n$ (See Proposition 6.7).\\

\noindent
{\bf PROPOSITION 2.12}: ([20,25]) The following recipe (already used implicitly in the Kalman test) will allow to bring an involutive system of order $q$ to an equivalent (isomorphic modules) involutive system of order 1 {\it with no zero order equations} called {\it Spencer form}:\\
1) Use a maximum set of arbitrary parametric derivatives up to order $q$ as new unknowns.\\
2) Make one prolongation.\\
3) Substitute the new unknowns.\\

\noindent
3)  {\bf GROUP MOTIVATION}:\\

   This section, which is a summary of results already obtained in [6], is provided for fixing the notations and the techniques leading to various (linear) {\it differential sequences}. All the results presented are local ones. A corresponding non-linear framework does exist but is out of the scope of this paper [6, 23].\\
Let $X$ be a manifold of dimension $n$ with local coordinates $x=(x^1,...,x^n)$ and latin indices $i,j=1,...,n$. We denote by $T=T(X)$ the tangent bundle to $X$ and by $T^*=T^*(X)$ the cotangent bundle to $X$ while ${\wedge}^rT^*$ is the bundle of $r$-forms on $X$. Also, we denote by $J_q(T)$ the $q$-{\it jet bundle} of $T$, that is to say the vector bundle over $X$ having the same transition rules as a vector field and its derivatives up to order $q$ under any change of local coordinates on $X$. Let now $G$ be a Lie group of dimension $p$ with identity $e$ and local coordinates $a=(a^{\tau})$ with $\tau=1,...,p$. We denote by ${\cal{G}}=T_{e}(G)$ the corresponding Lie algebra with vectors denoted by the greek letters $\lambda$. We shall identify a map $a:X\rightarrow G$, called a {\it gauging} of $G$ over $X$, with its graph $X\rightarrow X\times G$. We shall use the same notation for a bundle and its set of (local) sections as the background will always tell the right choice. In particular, when differential operators are involved, the sectional point of view must automatically be used. Such a convention allows to greatly simplify the notations at the expense of a slight abuse of language.\\

\noindent
{\bf DEFINITION 3.1}: A {\it Lie group of transformations} of a manifold $X$ is a lie group $G$ with an {\it action} of $G$ on $X$ better defined by its graph $X\times G \rightarrow X\times X: (x,a)\rightarrow (x,y=ax=f(x,a))$ with the properties that $a(bx)=(ab)x$ and $ex=x, \forall x\in X, \forall a,b\in G$. The action is {\it effective} if $ax=x, \forall x\in X\Rightarrow a=e$.\\

    Such groups of transformations have first been studied by S. Lie in 1880. Among basic examples when $n=1$ we may quote the {\it affine group} $y=ax+b$ and the {\it projective group} $y=(ax+b)/(cx+d)$ of transformations of the real line. When $n=3$ we may quote the {\it group of rigid motions} $y=Ax+B$ where now $A$ is an orthogonal $3\times 3$ matrix and $B$ is a vector. Only ten years later, in 1890,  Lie discovered that the Lie groups of transformations were only examples of  a wider class of groups of transformations, first called {\it infinite groups} but now called {\it Lie pseudogroups}.\\

\noindent
{\bf DEFINITION 3.2}: A Lie pseudogroup $\Gamma$ of transformations of a manifold $X$ is a group of transformations $y=f(x)$ solutions of a (in general nonlinear) system of OD/PD equations, also called system of {\it finite Lie equations}. For example, $y = a x + b$ is the generic solution of $y_{xx} = 0$.  \\

Setting now $y=x+t{\xi}(x)+...$ and passing to the limit for $t\rightarrow 0$, that is to say linearizing the defining system of finite Lie equations around the identity $y=x$, we get a linear system $R_q\subset J_q(T)$ for vector fields, also called system of {\it infinitesimal Lie equations}, with solutions $\Theta\subset T$ satisfying $[\Theta,\Theta]\subset \Theta$ and the corresponding operator on vector fields is called a {\it Lie operator}. It can be proved that such a system may be endowed with a Lie algebra bracket {\it on sections} ${\xi}_q:(x)\rightarrow (x,{\xi}^k(x), {\xi}^k_i(x),{\xi}^k_{ij}(x),...)$ as follows (See [1,2] for more details). Let us first define by bilinearity $\{j_{q+1}(\xi),j_{q+1}(\eta)\}=j_q([\xi,\eta]), \forall \xi,\eta\in T$ with $j_q(\xi):(x)\rightarrow (x,{\xi}^k(x), {\partial}_i{\xi}^k(x),{\partial}_{ij}{\xi}^k(x),...)$. Introducing the {\it Spencer operator} $d:R_{q+1}\rightarrow T^*\otimes R_q:{\xi}_{q+1}\rightarrow j_1({\xi}_q)-{\xi}_{q+1}$ with local components $({\partial}_i{\xi}^k-{\xi}^k_i,{\partial}_i{\xi}^k_j-{\xi}^k_{ij},...)$, we may set:\\
\[   [{\xi}_q,{\eta}_q]=\{{\xi}_{q+1},{\eta}_{q+1}\}+i(\xi)D{\eta}_{q+1}-i(\eta)D{\xi}_{q+1}  ,\forall {\xi}_q,{\eta}_q\in R_q\]
where $i( )$ is the interior multiplication (contraction) of a 1-form by a vector and we let the reader check that such a definition no longer depends on the "lifts" ${\xi}_{q+1},{\eta}_{q+1}$ over ${\xi}_q,{\eta}_q$. Such a bracket on sections transforms $R_q$ into a {\it Lie algebroid} in the sense that we have $[R_q,R_q]\subset R_q$ with $[{\xi}_q,{\eta}_q]=-[{\eta}_q,{\xi}_q]$ and the {\it Jacobi identity} 
$[[{\xi}_q,{\eta}_q],{\zeta}_q]+[[{\eta}_q,{\zeta}_q],{\xi}_q]+[[{\zeta}_q,{\xi}_q],{\eta}_q]=0 , \forall {\xi}_q,{\eta}_q,{\zeta}_q\in R_q$.\\

\noindent
{\bf EXAMPLE 3.3}: (Affine transformations)  $n=1,q=2, X=\mathbb{R}$\\
With evident notations, the system of finite Lie equations is defined by the single second order linear OD equation $y_{xx}=0$. Similarly, the {\it solutions} of $R_2$ are defined by ${\partial}_{xx}\xi(x)=0$ while the {\it sections} of $R_2$ are defined by ${\xi}_{xx}(x)=0$. Accordingly, the components of $[{\xi}_2,{\eta}_2]$ at order zero, one and two are defined by the totally unusual successive formulas:\\
\[    [\xi,\eta]=\xi{\partial}_x\eta-\eta{\partial}_x\xi     \]
\[    ([{\xi}_1,{\eta}_1])_x=\xi{\partial}_x{\eta}_x-\eta{\partial}_x{\xi}_x    \]
\[    ([{\xi}_2,{\eta}_2])_{xx}={\xi}_x{\eta}_{xx}-{\eta}_x{\xi}_{xx}+\xi{\partial}_x{\eta}_{xx}-\eta{\partial}_x{\xi}_{xx}   \]
It follows that ${\xi}_{xx}=0,{\eta}_{xx}=0\Rightarrow ([{\xi}_2,{\eta}_2])_{xx}=0$ and thus $[R_2,R_2]\subset R_2$.\\

In this apparently totally different framework, using the theorems of Lie, any action is locally generated by linearly independent infinitesimal generators $({\theta}_{\tau} = {\theta}^i_{\tau}(x) {\partial}_i \mid \tau:1,...,p)$ such that $[{\theta}_{\rho}, {\theta}_{\sigma} ] =c^{\tau}_{\rho \sigma} {\theta}^{\tau}$. If $J_q(T)$ is the $q$-jet bundle of $T$, we may introduce an operator $j_q:T \rightarrow J_q(T):{\xi}^k(x) \rightarrow ({\xi}^k(x), {\partial}_i {\xi}^k(x), {\partial}_{ij}{\xi}^k(x), ...) $ and so on up to o{rder $q$ included. Considering any section ${\xi}_q \in J_q(T)$ of the form ${\xi}^k_{\mu}(x) = {\lambda}^{\tau}(x) {\partial}_{\mu} {\theta}_{\tau}$, we obtain the first order Spencer operator $d:J_{q+1} \rightarrow T^* \otimes J_q(T)$ or simply $d {\xi}_{q+1}=j_1({\xi}_q)- {\xi}_{q+1}$ by the formula:\\
\[ \fbox{  $ (d {\xi}_{q+1})^k_{\mu,i}(x)   =    {\partial}_i{\xi}^k_{\mu}(x) - {\xi}^k_{\mu + 1_i}(x)  =  {\partial}_i {\lambda}^{\tau}(x) {\partial}_{\mu} {\theta}^k_{\tau}(x)    $  }  \]
that has never ben used for applications, in particular to control theory.  \\

    Introducing a basis of ${\wedge}^rT^*$ made by the $dx^I=dx^{i_1}\wedge ... \wedge dx^{i_r}$ with $I=(i_1<...<i_r)$, we may define the {\it exterior derivative} 
 $d:{\wedge}^rT^*\rightarrow {\wedge}^{r+1}T^*$ by setting $\omega={\omega}_Idx^I  \rightarrow d\omega ={\partial}_i{\omega}_Idx^i\wedge dx^I$ and one easily checks $d^2=d\circ d=0$. The (canonical linear) {\it gauge sequence}:\\
\[ \fbox{ $   {\wedge}^0T^*\otimes {\cal{G}}\stackrel{d}{\rightarrow}{\wedge}^1T^*\otimes {\cal{G}}\stackrel{d}{\rightarrow}{\wedge}^2T^*\otimes {\cal{G} } \stackrel{d}{\rightarrow}... \stackrel{d}{\rightarrow}{\wedge}^nT^*\otimes {\cal{G}}\rightarrow 0  $  }    \]
can be described by $p$ copies (indexed by $\tau$) of the Poincar\'{e} sequence for the exterior derivative.\\

    However, we did not speak about the other differential sequences that can be found in the literature, namely the {\it Janet sequence}, which is for sure the best known differential sequence, and the {\it Spencer sequence}. For short, starting from a vector bundle $E$ (for example $T$) and a linear differential operator ${\cal{D}}:E\rightarrow F:\xi\rightarrow \eta$ of order $q$, if we want to solve the linear system with second member ${\cal{D}}\xi=\eta$ even locally, one needs "{\it compatibility conditions}" (CC) in the form ${\cal{D}}_1\eta=0$. Denoting now $F$ by $F_0$, we may therefore look for an operator ${\cal{D}}_1:F_0\rightarrow F_1:\eta\rightarrow \zeta$ and so on. Under the assumption that $\cal{D}$ is involutive while taking into account the work of M. Janet in 1920 [21], one can prove that such a chain of operators ends after $n$ steps and we obtain the (canonical linear) {\it Janet sequence}, namely [1, 2]:\\
 \[  \fbox{  $  0\longrightarrow \Theta\longrightarrow E\stackrel{\cal{D}}{\longrightarrow}F_0\stackrel{{\cal{D}}_1}{\longrightarrow}F_1\stackrel{{\cal{D}}_2}{\longrightarrow}...
 \stackrel{{\cal{D}}_n}{\longrightarrow} F_n \longrightarrow 0   $  } \]
 where ${\cal{D}}_1,...,{\cal{D}}_n$ are first order involutive operators. The (canonical linear) {\it Spencer sequence} is the Janet sequence for the corresponding first order Spencer form, namely:\\
 \[  \fbox{  $   0\longrightarrow \Theta \stackrel{j_q}{\longrightarrow} C_0\stackrel{D_1}{\longrightarrow}C_1\stackrel{D_2}{\longrightarrow} ... \stackrel{D_n}{\longrightarrow}C_n\longrightarrow 0   $  }   \]
 where $C_0=R_q$ and the first order involutive operators $D_1,...,D_n$ are induced by the Spencer operator $D$ [1, 2]. It follows that, for any application where group theory is involved, we only have at our disposal the Janet sequence, the Spencer sequence and the gauge sequence. As these sequences are made by quite different operators, {\it the use of one excludes the use of the others}.\\
 
 In order to escape from this dilemna and for the sake of clarifying the key idea of the brothers Cosserat by using these new mathematical tools, we shall explain, in a way as elementary as possible in the linear framework, why {\it the Janet sequence and the gauge sequence cannot be used in continuum mechanics}. By this way we hope to convince the reader about the {\it need} to use another differential sequence, namely the Spencer sequence, though striking it could be. We notice that we have already exhibited the link existing between the gauge sequence and the Spencer sequence. Accordingly, the gauge sequence is isomorphic to the Spencer sequence:  \\
\[  \fbox{  $   0 \longrightarrow \Theta \stackrel{j_q}{\longrightarrow} {\wedge}^0T^*\otimes R_q\stackrel{D_1}{\longrightarrow} {\wedge}^1T^*\otimes R_q\stackrel{D_2}{\longrightarrow} {\wedge}^2T^*\otimes R_q \stackrel{D_3}{\longrightarrow} ... \stackrel{D_n}{\longrightarrow}{\wedge}^nT^*\otimes R_q\longrightarrow 0   $  }  \]
the isomorphisms being induced by the (local) isomorphism $X\times {\cal{G}}\rightarrow R_q$ of Lie algebroids just described above [1, 2]. It follows that {\it gauging} $\cal{G}$ {\it amounts to use an arbitrary section of} $R_q$ [24]. It is essential to notice that, though the Spencer sequence and the isomorphisms crucially depend on the action, by a kind of "{\it miracle}" the gauge sequence no longer depends on the action. Another difference lies in the fact that {\it all the indices} in the Spencer sequence range from $1$ to $n$ while in the gauge sequence the index $\tau$ ranges from $1$ to $p$. However, only the Spencer sequence can be used for Lie pseudogroups of transformations that are not coming from Lie groups of transformations. \\

Using the Stokes formula, the Cosserat  {\it couple- stress equations} are [17, 18]:  \\
\[        {\partial}_r{\sigma}^{ir}=f^i  , \hspace{2cm}  {\partial}_r{\mu}^{ij,r}+{\sigma}^{ij}-{\sigma}^{ji}=m^{ij}  \]
 This result shows that the surface density of forces $\vec{\sigma}$ and couples $\vec{\mu}$ is equivalent, from the point of view of torsor equilibrium, to a volume density of forces $\vec{f}$ and of momenta $\vec{m}$, providing the preceding stress and couple-stress equations are satisfied, and this interpretation explains the sign adopted. Of course, most of the engineering continua have the specific "{\it constitutive laws}"  $\mu=0, m=0$ and we get ${\sigma}^{ij}={\sigma}^{ji}$, a situation not always met in liquid crystals.\\

However, the combination of the stress {\it and} couple-stress equations have first been exhibited by E. and F. Cosserat in 1909 (See [5] and [16], p137,] {\it without any static equilibrium experimental background} and we now invite the reader to imagine how these equations could be related with the Spencer operator !.   \\
The following theorem leads to the same equations {\it just from group theoretical arguments}: \\

\noindent
{\bf THEOREM 3.4}: When $n=3, X={\mathbb{R}}^3$ and we deal with the group of rigid motions, the Cosserat equations are described by the formal adjoint of the first Spencer operator.   \\

\noindent
$\bullet$ {\it The gauge sequence cannot be used}: \\
Looking at the book [16] written by E. and F. Cosserat, it seems {\it at first sight} that they just construct the first operator of the gauge sequence for $n=1$ ( p 7), $n=2$ ( p 66), $n=3$ ( p 123) and finally $n=4$ ( p 189) in the linearized framework. {\it This is not true indeed} because the corresponding adjoint operator is a divergence like operator, a situation not met in the couple-stress equations. In fact, a carefull study of [16] proves that {\it somewhere in chapter 3 the action of the group on the space is used}, but this is well hidden among many very technical formulas and has never been noticed.   \\

\noindent
$\bullet $ {\it The Janet sequence cannot be used}:\\
This result is even more striking because {\it all} texbooks of elasticity use it along the same scheme that we now describe. Indeed, after gauging the translation by defining the "{\it displacement vector}"  $\xi=({\xi}^1(x),{\xi}^2(x))$ of the body, from the initial point $x=(x^1,x^2)$ to the point $y=x+\xi(x)$, one introduces the (small) "{\it deformation tensor}" $\epsilon=1/2{\cal{L}}(\xi)\omega$ as one half the Lie derivative with respect to $\xi$ of the euclidean metric $\omega$, namely, in our case, the three components only (care):\\
\[ \epsilon=({\epsilon}_{11}={\partial}_1{\xi}^1,{\epsilon}_{12}={\epsilon}_{21}=1/2({\partial}_1{\xi}^2+{\partial}_2{\xi}^1),{\epsilon}_{22}={\partial}_2{\xi}^2) = \frac{1}{2} \Omega    \]
One may check at once the only generating second order CC ${\cal{D}}_1\epsilon=0$, namely:\\
\[    {\partial}_{11}{\epsilon}_{22}+{\partial}_{22}{\epsilon}_{11}-2{\partial}_{12}{\epsilon}_{12}=0 \]
which is nothing else than the Riemann tensor of a metric, linearized at $\omega$.\\
For an arbitrary dimension $n$, one uses to consider the Lie operator ${\cal{D}}\xi={\cal{L}}(\xi)\omega:T\rightarrow S_2T^*$ (symmetric tensors), sometimes called {\it Killing operator} for the metric $\omega$, through the formula:\\
\[      ({\cal{D}}\xi)_{ij}\equiv {\omega}_{rj}{\partial}_i{\xi}^r+{\omega}_{ir}{\partial}_j{\xi}^r+{\xi}^r{\partial}_r{\omega}_{ij}={\Omega}_{ij} = {\Omega}_{ij} = 2{\epsilon}_{ij}   \]
and one obtains the $n^2(n^2-1)/12$ second order CC by linearizing at $\omega$ the Riemann tensor. 
However, the main experimental reason for introducing the first operator of this type of Janet sequence is the fact that the deformation is made from the displacement and first derivatives but must be invariant under any rigid motion. In the general case it must therefore have $(n+n^2)-(n+n(n-1)/2)=n(n+1)/2$ components, that is 3 when $n=2$, and this is the reason why introducing the deformation tensor $\epsilon$. For most finite element computations, the action density (local free energy) $w$ is a (in general quadratic) function of $\epsilon$ and people use to define the stress by the formula ${\sigma}^{ij}=\partial w/\partial {\epsilon}_{ij}$ which is not correct because $w$ only depends on ${\epsilon}_{11},{\epsilon}_{12}, {\epsilon}_{22}$ when $n=2$  as the deformation tensor is symmetric {\it by construction}. Finally, textbooks escape from this trouble by {\it deciding} that the stress should be symmetric and this is a {\it vicious circle} because we have proved it was not an assumption but an experimental result depending on specific constitutive laws. Accordingly, when $n=2$, we should have ${\sigma}^{ij}{\epsilon}_{ij} = {\sigma}^{11}{\epsilon}_{11}+(2{\sigma}^{12}){\epsilon}_{12}+{\sigma}^{22}{\epsilon}_{22}$. Hence, even if we find the correct stress equations with {\it this convenient duality} keeping the factor "2", {\it we have no way to get the stress and couple-stress equations together}.\\

\noindent
$ \bullet $ {\it Only the Spencer sequence can be used}:\\
Let us construct the formal adjoint of the Spencer operator by multiplying all the $(2\times 2)+2=6$ linearly independent nonzero components by corresponding test functions. For simplifying the summation, we shall raise and lower the indices by means of the (constant) euclidean metric, setting in particular ${\xi}_i={\omega}_{ir}{\xi}^r$ and ${\xi}_{i,j}={\omega}_{ir}{\xi}^r_j$. The only nonzero first jets coming from the $2\times 2$ skewsymmetric infinitesimal rotation matrix of first jets are now ${\xi}_{1,2}=-{\xi}_{2,1}$ while the second order jets are zero because isometries are linear transformations. We obtain the summation:\\
\[ \fbox{  $  {\sigma}^{11}{\partial}_1{\xi}_1+{\sigma}^{21}({\partial}_1{\xi}_2-{\xi}_{2,1})+{\sigma}^{12}({\partial}_2{\xi}_1-{\xi}_{1,2})+{\sigma}^{22}{\partial}_2{\xi}_2+{\mu}^{12,r}{\partial}_r{\xi}_{1,2}   $  }   \]
Integrating by parts and changing the sign, we just need to look at the coefficients of ${\xi}_1,{\xi}_2$ and ${\xi}_{1,2}$, namely:\\
\[  \left\{ \begin{array}{lclcl}
 {\xi}_1 & \longrightarrow  &  {\partial}_1{\sigma}^{11}+{\partial}_2{\sigma}^{12} & =  & f^1  \\
 {\xi}_2 & \longrightarrow   & {\partial}_1{\sigma}^{21}+{\partial}_2{\sigma}^{22} & = & f^2  \\
 {\xi}_{1,2} & \longrightarrow  & {\partial}_r{\mu}^{12,r}+{\sigma}^{12}-{\sigma}^{21} &= & m^{12}  
\end{array} \right. \]
in order to get the adjoint operator $ad(d):{\wedge}^{n-1}T^*\otimes R_1^*\rightarrow {\wedge}^nT^*\otimes R_1^*:(\sigma,\mu)\rightarrow (f,m)$ relating {\it for the first time} the torsor framework to the dual $R_1^*$ of the Lie algebroid $R_1$. These equations are {\it exactly} the three stress and couple-stress equations of 2-dimensional Cosserat elasticity. \\
For an arbitrary dimension $n$, the sections of $R_2$ satisfy ${\xi}_{i,j}+{\xi}_{j,i}=0, {\xi}^k_{ij}=0$ and we have to consider now the summation:\\
\[   {\sigma}^{ij}({\partial}_j{\xi}_i-{\xi}_{i,j})+{\sum}_{i<j}{\mu}^{ij,r}({\partial}_r{\xi}_{i,j})  \]
Integrating by part and changing the sign, we get, up to a divergence:\\
\[   {\partial}_r{\sigma}^{ir}{\xi}_i+{\sum}_{i<j}({\partial}_r{\mu}^{ij,r}+{\sigma}^{ij}-{\sigma}^{ji}){\xi}_{i,j}  \]
and obtain the {\it generalized Cosserat equations}:  \\
\[   \fbox{  $   {\partial}_r{\sigma}^{ir}=f^i, {\partial}_r{\mu}^{ij,r}+{\sigma}^{ij}-{\sigma}^{ji}=m^{ij}   $   }  \]
which can be used for the Poincar\' {e} group of space-time, even though, in this case, no direct approach can be provided.\\
Enlarging the group, the case of the conformal group of space-time could be treated similarly and the sections of the corresponding new system ${\hat{R}}_2\simeq{\hat{R}}_3$ satisfy: \\
\[  {\omega}_{rj}{\xi}^r_i+{\omega}_{ir}{\xi}^r_j-\frac{2}{n}{\omega}_{ij}{\xi}^r_r=0, n{\xi}^k_{ij}-{\delta}^k_i{\xi}^r_{rj}-{\delta}^k_j{\xi}^r_{ri}+{\omega}_{ij}{\omega}^{ks}{\xi}^r_{rs}=0, {\xi}^k_{ijr}=0, \forall n\geq 3   \]
where ${\delta}^k_i=0$ if $k\neq i$ or $1$ if $k=i$. Accordingly, among the components of the Spencer operator one may find ${\partial}_r{\xi}^k_{ij}-{\xi}^k_{ijr}={\partial}_r{\xi}^k_{ij}$ and thus the components 
${\partial}_i{\xi}^r_{rj}-{\partial}_j{\xi}^r_{ri}=F_{ij}$ of the EM field with EM potential ${\xi}^r_{ri}=A_i$ coming from the 4 elations, along lines only sketched by H. Weyl [25] because the needed mathematics were not available before 1970. Roughly, E. and F. Cosserat were only dealing with $\{{\xi}^k,{\xi}^k_i\mid {\xi}^r_r=0\}$ while, {\it in a somehow complementary way}, H. Weyl was only dealing with $\{{\xi}^r_r,{\xi}^r_{ri}\}$. The {\it new} Cosserat equation ${\xi}^r_r\rightarrow {\partial}_r{\mu}^r+ trace(\sigma)=0$ (in vacuum) explains why the trace of the EM enegy-momentum tensor vanishes as a consequence of the conservation of the density $\vec{\mu}$ of electric current ([25], \S 35, (74)) and the Spencer operator $D_2: T^*\otimes {\hat{R}}_2\rightarrow {\wedge}^2T^*\otimes {\hat{R}}_2$ ({\it field equations}) projects onto $d:{\wedge}^2T^*\rightarrow {\wedge}^3T^*$ ({\it Maxwell equations}). Such a result perfectly agrees with {\it piezzoelectricity} (quadratic lagrangian) and {\it photoelesticity} (cubic lagrangian) but {\it could not be obtained with the gauge sequence} and thus disagrees with {\it gauge theory} and the use of $U(1)$ [26].\\  \\  \\

\noindent
4)  {\bf PARAMETRIZATION PROBLEM}  \\

    The main tool in this section will be {\it duality theory}, namely the systematic use of the {\it formal adjoint} of an operator (see [6] for more details). For this, if $E$ is a vector bundle, we introduce its dual $E^*$ to be the vector bundle with inverse transition matrix (for example $T^*$ is the dual of $T$). The formal adjoint of an operator ${\cal{D}}:E\rightarrow F$ is the operator ${ad(\cal{D})}:{\wedge}^nT^*\otimes F^*\rightarrow {\wedge}^nT^*\otimes E^*$ defined by the following relation between volume forms:\\
\[    <\lambda,{\cal{D}}\xi>=<{ad(\cal{D})}\lambda,\xi> + d\alpha     \]
where $< >$ is the usual contraction, $\lambda$ is a test row vector density and $\alpha\in {\wedge} ^{n-1}T^*$ comes from Stokes formula of integration by part. Any operator can be considered as the formal adjoint of another operator because we have the identity $ad(ad(\cal{D}))=\cal{D}$. Also, if $P, Q\in D$ and $P=a^{\mu}d_{\mu}$, then $ad(P)=\sum (-1)^{\mid \mu \mid}d_{\mu}a^{\mu}$ and $ad(PQ)=ad(Q)ad(P)$.\\

Let us start with a given linear differential operator $\eta \stackrel{{\cal{D}}_1}{\longrightarrow } \zeta$ between the sections of two given vector bundles $F_0$ and $F_1$ of respective fiber dimension $m$ and $p$. Multiplying the equations ${\cal{D}}_1 \eta = \zeta$ by $p$ test functions $\lambda$ considered as a section of the adjoint vector bundle $ad(F_1)= {\wedge}^nT^*\otimes F^*_1$ and integrating by parts as we did in the introduction, we may introduce the adjoint vector bundle $ad(F_0)={\wedge}^nT^*\otimes F^*_0$ with sections $\mu$ in order to obtain the adjoint operator $ \mu \stackrel{ad({\cal{D}}_1)}{\longleftarrow} \lambda $, writing on purpose the arrow backwards, that is from right to left. More generally, let us consider a differential sequence:  \\
\[    \xi   \stackrel{\cal{D}}{\longrightarrow} \eta \stackrel{{\cal{D}}_1}{\longrightarrow} \zeta  \]
such that ${\cal{D}}_1$ generates the CC of ${\cal{D}}$ or, equivalently, such that ${\cal{D}}_1$ is parametrized by ${\cal{D}}$. \\
We may introduce the adjoint differential sequence:   \\
\[    \nu   \stackrel{ ad({\cal{D}} )}{\longleftarrow} \mu \stackrel{ ad( {\cal{D}}_1)}{\longleftarrow} \lambda  \]
As we have $ {\cal{D}}_1 \circ {\cal{D}}=0$, we obtain $ ad({\cal{D}}) \circ ad( {\cal{D}}_1) =0$. However, if ${\cal{D}}_1$ generates the CC of ${\cal{D}}$, then $ad({\cal{D}})$ may not generate the CC of $ad({\cal{D}}_1)$. Such a situation may not be satisfied as we shall see and the so-called {\it extension modules} have been introduced in order to measure these 
"{\it gaps} ". \\

The following nontrivial theorem, first obtained in [3], provides a {\it purely formal test} for deciding about the existence of a parametrization and exhibiting one. It is already implemented on the computer algebra package http://wwwb.math.rwth-aachen.de/OreModules.\\

\noindent
{\bf THEOREM 4.1}: Test for checking that a given differential module $M_1$ is torsion-free:  \\
$\fbox{1}$ Write the  corresponding defining operator ${\cal{D}}_1$.\\
$\fbox{2}$ Construct its formal adjoint $ad({\cal{D}}_1)$.\\
$\fbox{3}$  Work out generating CC for $ad({\cal{D}}_1)$ as an operator $ad({\cal{D}})$.\\
$\fbox{4}$  Construct $ad(ad({\cal{D}}))={\cal{D}}$.\\
$\fbox{5}$  Work out generating CC for ${\cal{D}}$ as an operator ${\cal{D}}'_1$.\\  
Then $M_1$ is torsion-free if and only if ${\cal{D}}_1$ and ${\cal{D}}'_1$ have the same solutions (both provide $M_1$).\\

 \[ \fbox{  $   \begin{array}{rcccccccl}
   &  &  &  &  &  &  &  &  \\
   &  &  &  & & & &  {\zeta}' &    \hspace{1cm}   \fbox{5}   \\
   &  &  &  & & &  \stackrel{ {\cal{D}}'_1}{\nearrow}  & &  \\
  \fbox{4} \hspace{7mm}  &  &  & \xi  & \stackrel{{\cal{D}}}{\longrightarrow} &  \eta & \stackrel{{\cal{D}}_1}{\longrightarrow} & \zeta         &\hspace{1cm}   \fbox{1}  \\
   &  &  &  &  &  &  &  &  \\
   & &  &  &  &  &  &   &  \\
  \fbox{3} \hspace{7mm}  &  &  & \nu & \stackrel{ad({\cal{D}})}{\longleftarrow} & \mu & \stackrel{ad({\cal{D}}_1)}{\longleftarrow} & \lambda &\hspace{1cm} \fbox{2}    \\ 
  & & & & & & & &     
  \end{array}  $  }    \] 
We have used the fact that $ad(ad({\cal{D}}))={\cal{D}}$ and that $ad({\cal{D}}) \circ ad({\cal{D}}_1)=0  \Rightarrow {\cal{D}}_1 \circ {\cal{D}}=0$, that is ${\cal{D}}_1$ is surely among the CC of ${\cal{D}}$ but other CC may also exist. \\

\noindent
{\bf COROLLARY 4.2}: Each new CC brought by ${\cal{D}}'_1$ which is not already a differential consequence of ${\cal{D}}_1$ is providing a torsion element of the differential module $M_1$ determined by ${\cal{D}}_1$. Hence ${\cal{D}}$ provides a parametrization of the system determined by $M_1/t(M_1)$ or, equivalently, $M'_1=M_1/t(M_1)$ is the torsion-free module determined by ${\cal{D}}'_1$ which is exactly the {\it minimum controllable realization} in classical control theory as we shall see.   \\

\noindent
{\bf COROLLARY 4.3}: When ${\cal{D}}_1$ can be parametrized, that is ${\cal{D}}_1$ constructed as in the theorem generates the CC of ${\cal{D}}$ or, equivalently, when $M_1$ is torsion-free and can be thus embedded into a free module $D^l$, we have thus $rk_D(M_1) = l'  \leq l $. There is a constructive procedure in order to embed $M_1$ into $D^{l'}$, that is to obtain a minimum parametrization. \\

The procedure with $4$ steps is as follows in the operator language (See Example 1.3):  \\
\noindent
$\fbox{1}$   Start with the formally exact {\it parametrizing sequence} already constructed by differential biduality.  We have thus $im({\cal{D}})=ker({\cal{D}}_1)$ and the corresponding diffferential module $M_1$ defined by ${\cal{D}}_1$ is torsion-free by assumption.  \\
$\fbox{2}$   Construct the adjoint sequence which is also formally exact by assumption.  \\
$\fbox{3}$   Find a {\it maximum} set of differentially independent CC $ad({\cal{D}}'):\mu \rightarrow {\nu}' $ among the generating CC $ad({\cal{D}} ):\mu \rightarrow  \nu$ of 
$ad({\cal{D}}_1)$ in such a way that $im(ad({\cal{D}}'))$ is a maximum free differential submodule of $im(ad({\cal{D}}))$ that is any element in $im(ad({\cal{D}}))$ is differentially 
algebraic over $im(ad({\cal{D}}'))$. \\
$\fbox{4}$     Using differential duality, construct ${\cal{D}}'=ad(ad({\cal{D}}'))$.            \\
Then ${\cal{D}}'$ is a minimum parametrization of ${\cal{D}}_1$.  \\

  \[  \begin{array}{rcccccccl}
  \fbox{4} \hspace{1cm} &  &  & {\xi}' & & & &  &          \\
   &  &  &\uparrow  &\stackrel{{\cal{D}}'}{\searrow} & & &  &  \\
   &  &  & \xi  & \stackrel{{\cal{D}}}{\longrightarrow} &  \eta & \stackrel{{\cal{D}}_1}{\longrightarrow} & \zeta         &\hspace{1cm}   \fbox{1}  \\
   &  &  &  &  &  &  &  &  \\
   &  &  &  &  &  &  &  &  \\
  &  &  & \nu & \stackrel{ad({\cal{D}})}{\longleftarrow} & \mu & \stackrel{ad({\cal{D}}_1)}{\longleftarrow} & \lambda &\hspace{1cm} \fbox{2}    \\
    &  &  &  \uparrow  &  \stackrel{ad({\cal{D}}')}{\swarrow} &  &  &   \\
 \fbox{3} \hspace{1cm}   &  &  &  {\nu}'   &  &  &  &  & \\
    &  &  \swarrow & \uparrow &  &  &  &  &   \\
    &  0 &   & 0 &  &  &  &  &
  \end{array}    \]    \\

Using the fact that the Poincar\'{e} sequence for the exterior derivative is self-adjoint {\it up to sign} (for $n=3$ the adjoints of $grad, curl, div$ are respectively $div, curl, grad$ up to sign) and that the extension modules do not depend on the sequence used for their definition [6, 20], we have:\\

\noindent
{\bf COROLLARY 4.4}: In the case of a Lie group of transformations, the gauge sequence is self-adjoint up to sign and thus $ad({\cal{D}})$ generates the CC of $ad({\cal{D}}_1)$ in the adjoint of any sequence where ${\cal{D}}_1$ generates the CC of the Lie operator $\cal{D}$ while $ad(D_1)$ generates the CC of $ad(D_2)$ in the adjoint of the corresponding Spencer sequence.\\

\noindent
{\bf COUNTEREXAMPLE 4.5}: Whith $m=1,n=2,q=2$, let us consider formally the involutive operator ${\cal{D}}:y\rightarrow (d_{12}y=u^1,d_{22}y=u^2)$ with ${\cal{D}}_1:(u^1,u^2)\rightarrow (d_1u^2-d_2u^1)$. Then $ad({\cal{D}}):({\phi}^1,{\phi}^2)\rightarrow (d_{12}{\phi}^1+d_{22}{\phi}^2)$ does not generate the CC of $ad({\cal{D}}_1):\lambda\rightarrow (d_2\lambda={\phi}^1,-d_1\lambda={\phi}^2)$ which are generated by the divergence condition $d_1{\phi}^1+d_2{\phi}^2=0$.\\

\noindent
{\bf EXAMPLE 4.6}: As a first striking consequence that does not seem to have been noticed by mechanicians up till now, let us consider the situation of classical elasticity theory where $\cal{D}$ is the Killing operator for the euclidean metric, namely $\cal{D}\xi=\cal{L}(\xi)\omega$ and ${\cal{D}}_1$ the corresponding CC, namely the linearized Riemann curvature with $n^2(n^2-1)/12$ components.  According to the above corollary, in order to parametrize the stress equations, that is $ad(\cal{D})$, one just needs to compute $ad({\cal{D}}_1)$. For $n=2$, we get:\\
\[    \phi ({\partial}_{11}{\epsilon}_{22}+{\partial}_{22}{\epsilon}_{11}-2{\partial}_{12}{\epsilon}_{12})={\partial}_{22}\phi{\epsilon}_{11}-2{\partial}_{12}\phi{\epsilon}_{12}+{\partial}_{11}\phi{\epsilon}_{22}+{\partial}_1(...)+{\partial}_2(...)   \]
and recover the parametrization by means of the Airy function in a rather unexpected way. For an arbitrary dimension $n$, this result is coherent with Example 2.7 as we have indeed $n(n-1)/2\leq n^2(n^2-1)/12, \forall n\geq 2$, with equality only for $n=2$.\\
 
\noindent
{\bf EXAMPLE 4.7}: We finally treat the case of the Cosserat equations. In this case we have $C_r={\wedge}^rT^*\otimes R_1\simeq {\wedge}^rT^*\otimes \cal{G}$ with ${dim(\cal{G})}=n(n+1)/2=p$. As we have shown in the last section that the Cosserat equations were just $ad(D_1)$, according to the above corollary a {\it first order} parametrization is thus described by $ad(D_2)$ and needs $dim(C_2)=n^2(n^2-1)/4$ potentials. We provide the details when $n=2$ but we know at once that we must use 
$3$ potentials only.\\
   The Spencer operator $D_1$ is described by the equations:\\
\[  {\partial}_1{\xi}_1=A_{11}, {\partial}_1{\xi}_2-{\xi}_{2,1}=A_{21}, {\partial}_2{\xi}_1-{\xi}_{1,2}=A_{12}, {\partial}_2{\xi}_2=A_{22}, {\partial}_1{\xi}_{1,2}=B_1, {\partial}_2{\xi}_{1,2}=B_2  \]
because $R_1$ is defined by the equations  ${\xi}_{1,1}=0,{\xi}_{1,2}+{\xi}_{2,1}=0, {\xi}_{2,2}=0$. \\
   Accordingly, the 3 CC describing the Spencer operator $D_2$ are:\\
\[  {\partial}_1A_{12}-{\partial}_2A_{11}+B_1=0, {\partial}_1A_{22}-{\partial}_2A_{21}+B_2=0, 
{\partial}_1B_2-{\partial}_2B_1=0  \]
Multiplying these equations respectively by ${\phi}^1, {\phi}^2, {\phi}^3$, then summing and integrating by part, we get $ad(D_2)$ and the desired first order parametrization in the form:\\
\[  \fbox{  $  {\sigma}^{11}={\partial}_2{\phi}^1, {\sigma}^{12}=-{\partial}_1{\phi}^1, {\sigma}^{21}={\partial}_2{\phi}^2, {\sigma}^{22}=-{\partial}_1{\phi}^2, {\mu}^{12,1}={\partial}_2{\phi}^3+{\phi}^1, {\mu}^{12,2}=-{\partial}_1{\phi}^3+{\phi}^2  $  }  \]
as announced previously. It is important to notice that such a parametrization, which could also be obtained by localization, is coherent with the classical one already obtained by localization in Example 2.6, which can be recovered if we cancel the couple-stress and set ${\phi}^3=-\phi$. \\
For an arbitrary dimension $n$, $D_1$ is given by the $n^2(n+1)/2$ equations  
${\partial}_i{\xi}^k-{\xi}^k_i=A^k_i, {\partial}_i{\xi}^k_j=B^k_{j,i}$ and $D_2$ provides the $n^2(n^2-1)/4$ CC:\\
\[   {\partial}_iA^k_j-{\partial}_jA^k_i+B^k_{j,i}-B^k_{i,j}=0  ,  {\partial}_iB^k_{r,j}-{\partial}_jB^k_{r,i}=0  \]
Lowering the index $k$ and contracting them respectively by test functions ${\phi}^{k,ij}=-{\phi}^{k,ji}$ and ${\psi}^{kr,ij}=-{\psi}^{rk,ij}=-{\psi}^{kr,ji}$ for $i< j$, then integrating by part, we obtain the first order parametrization $ad(D_2)$:\\
\[  {\sigma}^{ij}={\partial}_r{\phi}^{i,jr} \hspace{1cm},\hspace{1cm} {\mu}^{ij,r}={\partial}_s{\psi}^{ij,rs}+{\phi}^{j,ir}-{\phi}^{i,jr}  \]
This result is coherent with the fact that the minimum number of potentials is now $(n-1)p$ as we have indeed $n(n^2-1)/2\leq n^2(n^2-1)/4, \forall n\geq 2$, with equality only for $n=2$. Using the conformal group of space-time provides a common parametrization for the Cosserat and Maxwell equations in a {\it unique framework}.\\

\noindent
{\bf 5) SYSTEM THEORY}  \\

For example, the fact that the Cauchy operator is the adjoint of the Killing operator for the Euclidean metric is in any textbook of continuum mechanics in the chapter "variational calculus" and the parametrization problem has been quoted by many famous authors, as we said in the Abstract, but only from a computational point of view. However it is still not known that the adjoint of the $20$ components of the Bianchi operator has been introduced by C. Lanczos as we explained with details in [28]. However, the main trouble is that these two problems have {\it never} ben treated in an intrinsic way and, in particular, changes of coordinates have {\it never} been considered. The same situation can be met for Maxwell equations but is out of our scope [26].   \\

\noindent
{\bf PROPOSITION 5.1}: The Cauchy operator is the adjoint of the Killing operator in arbitrary dimension, up to sign.  \\

\noindent
{\it Proof}: Let $X$ be a manifold of dimension $n$ with local coordinates $(x^1, ... , x^n)$, tangent bundle $T$ and cotangent bundle $T^*$. If $\omega \in S_2 T^*$ is a metric with $det(\omega)\neq 0$, we my introduce the standard Lie derivative in order to define the first order Killing operator:    \\
\[       {\cal{D}}: \xi \in T \rightarrow \Omega = 
({\Omega}_{ij}= {\omega}_{rj}(x) {\partial}_i{\xi}^r + {\omega}_{ir}(x){\partial}_j{\xi}^r + 
{\xi}^r {\partial}_r {\omega}_{ij}(x) ) \in S_2T^*  \]
Here start the problems because, in our opinion at least, a systematic use of the adjoint operator has never been used in mathematical physics and even in continuum mechanics apart through a variational procedure. As will be seen later on, the purely intrinsic definition of the adjoint can only be done in the theory of differential modules by means of the so-called {\it side changing functor}. From a purely differential geometric point of view, the idea is to associate to any vector bundle $E$ over $X$ a new vector bundle 
$ad(E)= {\wedge}^nT^* \otimes E^*$ where $E^*$ is obtained from $E$ by patching local coordinates while inverting the transition matrices, exactly like $T^*$ is obtained from $T$. It follows that the stress tensor $\sigma =({\sigma}^{ij}) \in ad(S_2T^*) =  {\wedge}^n T^* \otimes S_2 T$ is {\it not} a tensor but a tensor density, that is transforms like a tensor up to a certain power of the Jacobian matrix. When $n=4$,  the fact that such an object is called stress-energy tensor does not change anything as it cannot be related to the Einstein tensor which is a true {\it tensor} indeed. Of course, it is always possible in GR to use $(det(\omega))^\frac{1}{2} $ but, as we shall see, the study of contact structures {\it must} be done without any reference to a background metric. In any case, using the metric to raise or lower the indices, we may define:  \\
\[  ad({\cal{D}}): {\wedge}^n T^* \otimes S_2T \rightarrow {\wedge}^n T^* \otimes T : \sigma \rightarrow \varphi \]
Multiplying ${\Omega}_{ij}$ by ${\sigma}^{ij}$ and integrating by parts, the factor of $ - 2 {\xi}^k$ is easly seen to be:  \\
\[   {\nabla}_i {\sigma}^{ik} = {\partial}_i {\sigma}^{ik} +  {\gamma}^k_{ij} {\sigma}^{ij} = {\varphi}^k   \] 
with well known Christoffel symbols 
$ {\gamma}^k_{ij} = \frac{1}{2} {\omega}^{kr} ({\partial}_i {\omega}_{rj} + {\partial}_j {\omega}_{ir} - 
{\partial}_r {\omega}_{ij}) $. \\
However, {\it if the stress should be a tensor}, we should get for the covariant derivative:  \\
\[ {\nabla}_r {\sigma}^{ij}= {\partial}_r{\sigma}^{ij} + {\gamma}^i_{rs}{\sigma}^{sj}+ {\gamma}^j_{rs} {\sigma}^{is} \Rightarrow 
{\nabla}_i {\sigma}^{ik} = {\partial}_i {\sigma}^{ik} + {\gamma}^r_{ri} {\sigma}^{ik} + 
{\gamma}^k_{ij} {\sigma}^{ij}    \]
The difficulty is to prove that we do not have a contradiction because $\sigma$ is a tensor density. This tricky technical result, which is not evident at all, explains why the additional term we had is just disappearing in fact when $\sigma$ is a density. \\
\hspace*{12cm}  $   \Box   $   \\

If $X$ is a manifold of dimension $n$ with local coordinates $(x)=(x^1, ... ,x^n)$, we denote as usual by $T=T(X)$ the {\it tangent bundle} of $X$, by $T^*=T^*(X)$ the {\it cotangent bundle}, by ${\wedge}^rT^*$ the {\it bundle of r-forms} and by $S_qT^*$ the {\it bundle of q-symmetric tensors}. More generally, let $E$ be a {\it vector bundle} over $X$ with local coordinates $(x^i,y^k)$ for $i=1,...,n$ and $k=1,...,m$ simply denoted by $(x,y)$, {\it projection} $\pi:E\rightarrow X:(x,y)\rightarrow (x)$ and changes of local coordinate $\bar{x}=\varphi (x), \bar{y}=A(x)y$. We shall denote by $E^*$ the vector bundle obtained by inverting the matrix $A$ of the changes of coordinates, exactly like $T^*$ is obtained from $T$. We denote by $f:X\rightarrow E: (x)\rightarrow (x,y=f(x))$ a global {\it section} of $E$, that is a map such that $\pi\circ f=id_X$ but local sections over an open set $U\subset X$ may also be considered when needed. Under a change of coordinates, a section transforms like $\bar{f}(\varphi(x))=A(x)f(x)$ and the changes of the derivatives can also be obtained with more work. We shall denote by $J_q(E)$ the {\it q-jet bundle} of $E$ with local coordinates $(x^i, y^k, y^k_i, y^k_{ij},...)=(x,y_q)$ called {\it jet coordinates} and sections $f_q:(x)\rightarrow (x,f^k(x), f^k_i(x), f^k_{ij}(x), ...)=(x,f_q(x))$ transforming like the sections $j_q(f):(x) \rightarrow (x,f^k(x), {\partial}_if^k(x), {\partial}_{ij}f^k(x), ...)=(x,j_q(f)(x))$ where both $f_q$ and $j_q(f)$ are over the section $f$ of $E$. For any $q\geq 0$, $J_q(E)$ is a vector bundle over $X$ with projection ${\pi}_q$ while $J_{q+r}(E)$ is a vector bundle over $J_q(E)$ with projection ${\pi}^{q+r}_q, \forall r\geq 0$.\\

\noindent
{\bf DEFINITION  5.2}: A {\it linear system} of order $q$ on $E$ is a vector sub-bundle $R_q\subset J_q(E)$ and a {\it solution} of $R_q$ is a section $f$ of $E$ such that $j_q(f)$ is a section of $R_q$. With a slight abuse of language, the set of local solutions will be denoted by $\Theta \subset E$.\\
  
Let $\mu=({\mu}_1,...,{\mu}_n)$ be a multi-index with {\it length} ${\mid}\mu{\mid}={\mu}_1+...+{\mu}_n$, {\it class} $i$ if ${\mu}_1=...={\mu}_{i-1}=0,{\mu}_i\neq 0$ and $\mu +1_i=({\mu}_1,...,{\mu}_{i-1},{\mu}_i +1, {\mu}_{i+1},...,{\mu}_n)$. We set $y_q=\{y^k_{\mu}{\mid} 1\leq k\leq m, 0\leq {\mid}\mu{\mid}\leq q\}$ with $y^k_{\mu}=y^k$ when ${\mid}\mu{\mid}=0$. If $E$ is a vector bundle over $X$ and $J_q(E)$ is the $q$-{\it jet bundle} of $E$, then both sections $f_q\in J_q(E)$ and $j_q(f)\in J_q(E)$ are over the section $f\in E$. There is a natural way to distinguish them by introducing the {\it Spencer}  operator $d:J_{q+1}(E)\rightarrow T^*\otimes J_q(E)$ with components $(df_{q+1})^k_{\mu,i}(x)={\partial}_if^k_{\mu}(x)-f^k_{\mu+1_i}(x)$. The kernel of $d$ consists of sections such that $f_{q+1}=j_1(f_q)=j_2(f_{q-1})=...=j_{q+1}(f)$. Finally, if $R_q\subset J_q(E)$ is a {\it system} of order $q$ on $E$ locally defined by linear equations ${\Phi}^{\tau}(x,y_q)\equiv a^{\tau\mu}_k(x)y^k_{\mu}=0$ and local coordinates $(x,z)$ for the parametric jets up to order $q$, the $r$-{\it prolongation} $R_{q+r}={\rho}_r(R_q)=J_r(R_q)\cap J_{q+r}(E)\subset J_r(J_q(E))$ is locally defined when $r=1$ by the linear equations ${\Phi}^{\tau}(x,y_q)=0, d_i{\Phi}^{\tau}(x,y_{q+1})\equiv a^{\tau\mu}_k(x)y^k_{\mu+1_i}+{\partial}_ia^{\tau\mu}_k(x)y^k_{\mu}=0$ and has {\it symbol} $g_{q+r}=R_{q+r}\cap S_{q+r}T^*\otimes E\subset J_{q+r}(E)$ if one looks at the {\it top order terms}. If $f_{q+1}\in R_{q+1}$ is over $f_q\in R_q$, differentiating the identity $a^{\tau\mu}_k(x)f^k_{\mu}(x)\equiv 0$ with respect to $x^i$ and substracting the identity $a^{\tau\mu}_k(x)f^k_{\mu+1_i}(x)+{\partial}_ia^{\tau\mu}_k(x)f^k_{\mu}(x)\equiv 0$, we obtain the identity $a^{\tau\mu}_k(x)({\partial}_if^k_{\mu}(x)-f^k_{\mu+1_i}(x))\equiv 0$ and thus the restriction $d:R_{q+1}\rightarrow T^*\otimes R_q$. More generally, we have the restriction:   \\
\[   d: {\wedge}^sT^* \otimes R_{q+1} \rightarrow {\wedge}^{s+1}T^* \otimes R_q: (f^k_{\mu,I}(x)dx^I) \rightarrow (({\partial}_if^k_{\mu,I}(x) - f^k_{\mu + 1_i,I}(x))dx^i \wedge dx^I)\] 
with standard multi-index notation for exterior forms and one can easily check that $d\circ d=0$. The restriction of $-d$ to the symbol is called the {\it Spencer} map $\delta$ in the sequences:  \\
\[ {\wedge}^{s-1}T^*\otimes g_{q+r + 1} \stackrel{\delta}{\longrightarrow} {\wedge}^s T^* \otimes g_{q+r} 
\stackrel{\delta}{\longrightarrow} {\wedge}^{s+1} T^* \otimes g_{q + r -1}  \]
because $\delta \circ \delta=0$ similarly, leading to the purely algebraic {\it $\delta$-cohomology} $H^s_{q+r}(g_q)$ at $ {\wedge}^sT^* \otimes g_{q+r}$ with similar notation for the coboundary $B = im (\delta)\subseteq  Z = ker( \delta) \Rightarrow H = Z/B$ bundles. \\

\noindent
{\bf DEFINITION 5.3}: A system $R_q$ is said to be {\it formally integrable} when all the equations of order $q+r$ are obtained by $r$ prolongations {\it only}, $\forall r\geq 0$ or, equivalently, when the projections ${\pi}^{q+r+s}_{q+r}:R_{q+r+s}\rightarrow R^{(s)}_{q+r} \subseteq R_{q+r}$ are such that $R^{(s)}_{q+r} = R_{q+r}$, $\forall r,s\geq0$.  \\

Finding an intrinsic test has been achieved by D.C. Spencer in 1970 [11] along coordinate dependent lines sketched by M. Janet in 1920 [21]. The next procedure providing a {\it Pommaret basis} and where  {\it one may have to change linearly the independent variables if necessary}, is intrinsic even though it must be checked in a particular coordinate system called $\delta$-{\it regular} [1, 6].  \\

$\bullet$ {\it Equations of class} $n$: Solve the maximum number ${\beta}^n_q$ of equations with respect to the jets of order $q$ and class $n$. Then call $(x^1,...,x^n)$ {\it multiplicative variables}.\\

$\bullet$ {\it Equations of class} $i\geq 1$: Solve the maximum number ${\beta}^i_q$ of {\it remaining} equations with respect to the jets of order $q$ and class $i$. Then call $(x^1,...,x^i)$ {\it multiplicative variables} and $(x^{i+1},...,x^n)$ {\it non-multiplicative variables}.\\

$\bullet$ {\it Remaining equations equations of order} $\leq q-1$: Call $(x^1,...,x^n)$ {\it non-multiplicative variables}.\\

\noindent
In actual practice, we shall use a {\it Janet tabular} where the multiplicative "variables" are in upper left position while the non-multiplicative variables are represented by dots in lower right position.  \\

\noindent
{\bf DEFINITION 5.4}: A system of PD equations is said to be {\it involutive} if its first prolongation can be obtained by prolonging its equations only with respect to the corresponding multiplicative variables. In that case, we may introduce the {\it characters} ${\alpha}^i_q=m\frac{(q+n-i-1)!}{(q-1)!((n-i)!}-{\beta}^i_q$ for $i=1, ..., n$ with ${\alpha}^1_q\geq ... \geq {\alpha}^n_q\geq 0$ and we have $dim(g_q)={\alpha}^1_q+...+ {\alpha}^n_q$ while $dim(g_{q+1})={\alpha}^1_q+...+ n {\alpha}^n_q$. \\

\noindent
{\bf REMARK 5.5}: As long as the {\it Prolongation}/{\it Projection} (PP) procedure allowing to find two integers $r,s\geq 0$ such that the system $R^{(s)}_{q+r}$ is involutive, has not been achieved, {\it nothing} can be said about the CC (Fine examples can be found in the recent [29]). \\

When $R_q$ is involutive, the operator ${\cal{D}}:E\stackrel{j_q}{\rightarrow} J_q(E)\stackrel{\Phi}{\rightarrow} J_q(E)/R_q=F_0$ of order $q$ is said to be {\it involutive}. Introducing the {\it Janet bundles} $F_r= {\wedge}^rT^*\otimes J_q(E)/({\wedge}^rT^*\otimes R_q + \delta (S_{q+1}T^*\otimes E))$, we obtain the {\it linear Janet sequence} (Introduced in [1, 2]):\\
\[ \fbox{  $   0 \longrightarrow  \Theta \longrightarrow E \stackrel{\cal{D}}{\longrightarrow} F_0 \stackrel{{\cal{D}}_1}{\longrightarrow}F_1 \stackrel{{\cal{D}}_2}{\longrightarrow} ... \stackrel{{\cal{D}}_n}{\longrightarrow} F_n \longrightarrow 0  $  }   \]
where each other operator is first order involutive and generates the CC of the preceding one. \\
Similarly, introducing the {\it Spencer bundles} $C_r= {\wedge}^rT^* \otimes R_q / \delta ({\wedge}^{r-1}T^* \otimes g_{q+1})$ we obtain the {\it linear Spencer sequence} induced by the Spencer operator [1, 2]:  \\
\[  \fbox{  $  0\longrightarrow \Theta \stackrel{j_q}{\longrightarrow} C_0 \stackrel{D_1}{\longrightarrow} C_1 \stackrel{D_2}{\longrightarrow}... \stackrel{D_n}{\longrightarrow} C_n \longrightarrow  0 $  }   \]   \\\

\vspace{2cm}

\noindent
{\bf 6) MODULE THEORY}:   \\

Let $K$ be a {\it differential field} with $n$ commuting derivations $({\partial}_1,...,{\partial}_n)$ and consider the ring $D=K[d_1,...,d_n]=K[d]$ of differential operators with coefficients in $K$ with $n$ commuting formal derivatives satisfying $d_ia=ad_i + {\partial}_ia$ in the operator sense. If $P=a^{\mu}d_{\mu}\in D=K[d]$, the highest value of ${\mid}\mu {\mid}$ with $a^{\mu}\neq 0$ is called the {\it order} of the {\it operator} $P$ and the ring $D$ with multiplication $(P,Q)\longrightarrow P\circ Q=PQ$ is filtred by the order $q$ of the operators. We have the {\it filtration} $0\subset K=D_0\subset D_1\subset  ... \subset D_q \subset ... \subset D_{\infty}=D$. As an algebra, $D$ is generated by $K=D_0$ and $T=D_1/D_0$ with $D_1=K\oplus T$ if we identify an element $\xi={\xi}^id_i\in T$ with the vector field $\xi={\xi}^i(x){\partial}_i$ of differential geometry, but with ${\xi}^i\in K$ now. It follows that $D={ }_DD_D$ is a {\it bimodule} over itself, being at the same time a left $D$-module by the composition $P \longrightarrow QP$ and a right $D$-module by the composition $P \longrightarrow PQ$. We define the {\it adjoint} functor $ad:D \longrightarrow D^{op}:P=a^{\mu}d_{\mu} \longrightarrow  ad(P)=(-1)^{\mid \mu \mid}d_{\mu}a^{\mu}$ and we have $ad(ad(P))=P$ both with $ad(PQ)=ad(Q)ad(P), \forall P,Q\in D$. Such a definition can be extended to any matrix of operators by using the transposed matrix of adjoint operators (See [6] for more details and applications to control theory or mathematical physics). \\

Accordingly, if $y=(y^1, ... ,y^m)$ are differential indeterminates, then $D$ acts on $y^k$ by setting $d_iy^k=y^k_i \longrightarrow d_{\mu}y^k=y^k_{\mu}$ with $d_iy^k_{\mu}=y^k_{\mu+1_i}$ and $y^k_0=y^k$. We may therefore use the jet coordinates in a formal way as in the previous section. Therefore, if a system of OD/PD equations is written in the form ${\Phi}^{\tau}\equiv a^{\tau\mu}_ky^k_{\mu}=0$ with coefficients $a\in K$, we may introduce the free differential module $Dy=Dy^1+ ... +Dy^m\simeq D^m$ and consider the differential {\it module of equations} $I=D\Phi\subset Dy$, both with the residual {\it differential module} $M=Dy/D\Phi$ or $D$-module and we may set $M={ }_DM$ if we want to specify the ring of differential operators. We may introduce the formal {\it prolongation} with respect to $d_i$ by setting $d_i{\Phi}^{\tau}\equiv a^{\tau\mu}_ky^k_{\mu+1_i}+({\partial}_ia^{\tau\mu}_k)y^k_{\mu}$ in order to induce maps $d_i:M \longrightarrow M:{\bar{y} }^k_{\mu} \longrightarrow {\bar{y}}^k_{\mu+1_i}$ by residue with respect to $I$ if we use to denote the residue $Dy \longrightarrow M: y^k \longrightarrow {\bar{y}}^k$ by a bar like in algebraic geometry. However, for simplicity, we shall not write down the bar when the background will indicate clearly if we are in $Dy$ or in $M$. As a byproduct, the differential modules we shall consider will always be {\it finitely generated} ($k=1,...,m<\infty$) and {\it finitely presented} ($\tau=1, ... ,p<\infty$). Equivalently, introducing the {\it matrix of operators} ${\cal{D}}=(a^{\tau\mu}_kd_{\mu})$ with $m$ columns and $p$ rows, we may introduce the morphism $D^p \stackrel{{\cal{D}}}{\longrightarrow} D^m:(P_{\tau}) \longrightarrow (P_{\tau}{\Phi}^{\tau})$ over $D$ by acting with $D$ {\it on the left of these row vectors} while acting with ${\cal{D}}$ {\it on the right of these row vectors} by composition of operators with $im({\cal{D}})=I$. The {\it presentation} of $M$ is defined by the exact cokernel sequence $D^p \stackrel{{\cal{D}}}{\longrightarrow} D^m \longrightarrow M \longrightarrow 0 $. We notice that the presentation only depends on $K, D$ and $\Phi$ or $ \cal{D}$, that is to say never refers to the concept of (explicit local or formal) solutions. It follows from its definition that $M$ can be endowed with a {\it quotient filtration} obtained from that of $D^m$ which is defined by the order of the jet coordinates $y_q$ in $D_qy$. We have therefore the {\it inductive limit} $0 \subseteq M_0 \subseteq M_1 \subseteq ... \subseteq M_q \subseteq ... \subseteq M_{\infty}=M$ with $d_iM_q\subseteq M_{q+1}$ and $M=DM_q$ for $q\gg 0$ with prolongations $D_rM_q\subseteq M_{q+r}, \forall q,r\geq 0$. It is important to notice that it may be sometimes quite difficult to work out $I_q$ or $M_q$ from a given presentation which is not involutive [29].   \\ 

\noindent
{\bf DEFINITION 6.1}: An exact sequence of morphisms finishing at $M$ is said to be a {\it resolution} of $M$. If the differential modules involved apart from $M$ are free, that is isomorphic to a certain power of $D$, we shall say that we have a {\it free resolution} of $M$.  \\

Having in mind that $K$ is a left $D$-module with the action $(D,K) \longrightarrow K:(d_i,a)\longrightarrow {\partial}_ia$ and that $D$ is a bimodule over itself with $PQ\neq QP$, {\it we have only two possible constructions}:  \\

\noindent
{\bf DEFINITION 6.2}: We may define the right ({\it care})  differential module $hom_D(M,D) $ with $(f P)(m) = (f(m)) P \Rightarrow (fPQ)(m)=((fP)(m))Q=((f(m))P)Q=(f(m))PQ$.  \\                        

\noindent
{\bf DEFINITION 6.3}: We define the {\it system} $R=hom_K(M,K)$ and set $R_q=hom_K(M_q,K)$ as the {\it system of order} $q$. We have the {\it projective limit} $R=R_{\infty} \longrightarrow ... \longrightarrow R_q \longrightarrow ... \longrightarrow R_1 \longrightarrow R_0$. It follows that $f_q\in R_q:y^k_{\mu} \longrightarrow f^k_{\mu}\in K$ with $a^{\tau\mu}_kf^k_{\mu}=0$ defines a {\it section at order} $q$ and we may set $f_{\infty}=f\in R$ for a {\it section} of $R$. For an arbitrary differential field $K$, {\it such a definition has nothing to do with the concept of a formal power series solution} ({\it care}).\\

\noindent
{\bf PROPOSITION 6.4}: When $M$ is a left $D$-module, then $R$ is also a left $D$-module. \\

\noindent
{\it Proof}: As $D$ is generated by $K$ and $T$ as we already said, let us define:  \\
\[  (af)(m)=af(m)= f(am), \hspace{4mm} \forall a\in K, \forall m\in M \]
\[ (\xi f)(m)=\xi f(m)-f(\xi m), \hspace{4mm} \forall \xi=a^id_i\in T,\forall m\in M  \]
In the operator sense, it is easy to check that $d_ia=ad_i+{\partial}_ia$ and that $\xi\eta - \eta\xi=[\xi,\eta]$ is the standard bracket of vector fields. We finally 
get $(d_if)^k_{\mu}=(d_if)(y^k_{\mu})={\partial}_if^k_{\mu}-f^k_{\mu +1_i}$ and thus recover {\it exactly} the Spencer operator of the previous section though {\it this is not evident at all}. We also get $(d_id_jf)^k_{\mu}={\partial}_{ij}f^k_{\mu}-{\partial}_if^k_{\mu+1_j}-{\partial}_jf^k_{\mu+1_i}+f^k_{\mu+1_i+1_j} \Longrightarrow d_id_j=d_jd_i, \forall i,j=1,...,n$ and thus $d_iR_{q+1}\subseteq R_q\Longrightarrow d_iR\subset R$ induces a well defined operator $R\longrightarrow T^*\otimes R:f \longrightarrow dx^i\otimes d_if$. This operator has been first introduced, up to sign, by F.S. Macaulay as early as in $1916$ but this is still not ackowledged [22]. For more details on the Spencer operator and its applications, the reader may look at [23].  \\
\hspace*{12cm}   $  \Box   $   \\

The two following definitions, which are well known in commutative algebra, are also valid (with more work) in the case of differential modules (See [6] for more details or the references [10, 19, 20] for an introduction to homological algebra and diagram chasing).  \\

\noindent
{\bf DEFINITION 6.5}: The set of elements $t(M) = \{ m \in M \mid \exists 0 \neq P \in D, Pm=0\}\subseteq M$ is a differential module called the {\it torsion submodule} of $M$. More generally, a module $M$ is called a {\it torsion module} if $t(M)=M$ and a {\it torsion-free module} if $t(M)=0$. In the short exact sequence $0 \rightarrow t(M) \rightarrow M \rightarrow M' \rightarrow 0$, the module $M'$ is torsion-free. Its defining module of equations $I'$ is obtained by adding to $I$ a representative basis of $t(M)$ set up to zero and we have thus $I \subseteq I'$.  \\

\noindent
{\bf DEFINITION 6.6}: A differential module $F$ is said to be {\it free} if $F \simeq D^r$ for some integer 
$r > 0$ and we shall {\it define} $rk_D(F)=r$. If $F$ is the biggest free differential module contained in $M$, then $M/F$ is a torsion differential module and $hom_D(M/F,D)=0$. In that case, we shall {\it define} the {\it differential rank} of $M$ to be $rk_D(M)=rk_D(F)=r$. Accordingly, if $M$ is defined by a linear involutive operator of order $q$, then $rk_D(M)={\alpha}^n_q$.  \\

\noindent
{\bf PROPOSITION 6.7}: If $0 \rightarrow M' \rightarrow M \rightarrow M" \rightarrow 0$ is a short exact sequence of differential modules and maps or operators, we have $rk_D(M)=rk_D(M') + rk_D(M")$.  \\

In the general situation, let us consider the sequence $ M' \stackrel{f}{\longrightarrow} M  \stackrel{g}{\longrightarrow} M" $ of modules which may not be exact and define $B =im(f) \subseteq Z = ker(g) \Rightarrow   H=Z/B$. \\

In order to conclude this section, we may say that the main difficulty met when passing from the differential framework to the algebraic framework is the " {\it inversion} " of arrows. Indeed, when an operator is injective, that is when we have the exact sequence $ 0 \rightarrow E \stackrel{{\cal{D}}}{\longrightarrow} F$ with $dim(E)=m, dim(F)=p$, like in the case of the operator $0 \rightarrow E \stackrel{j_q}{\longrightarrow} J_q(E) $, on the contrary, using differential modules, we have the epimorphism 
$D^p \stackrel{{\cal{D}}}{\longrightarrow} D^m \rightarrow 0$. The case of a formally surjective operator, like the $div$ operator, described by the exact sequence $E \stackrel{{\cal{D}}}{\longrightarrow} F \rightarrow 0$ is now providing the exact sequence of differential modules $ 0 \rightarrow D^p \stackrel{{\cal{D}}}{\longrightarrow} D^m \rightarrow M  \rightarrow 0$ because ${\cal{D}}$ has no CC.  \\   
In order to conclude this section, we may say that the main difficulty met when passing from the differential framework to the algebraic framework is the " {\it inversion} " of arrows. Indeed, with $dim(E)=m, dim(F)=p$, when an operator ${\cal{D}}:E \rightarrow F$ is injective, that is when we have the exact sequence $ 0 \rightarrow E \stackrel{{\cal{D}}}{\longrightarrow} F$, like in the case of the operator $0 \rightarrow E \stackrel{j_q}{\longrightarrow} J_q(E) $, on the contrary, using differenial modules, we have the epimorphism $D^p \stackrel{{\cal{D}}}{\longrightarrow} D^m \rightarrow 0$. The case of a formally surjective operator, like the $div$ operator, described by the exact sequence $E \stackrel{{\cal{D}}}{\longrightarrow} F \rightarrow 0$ is now providing the exact sequence of differential modules $ 0 \rightarrow D^p \stackrel{{\cal{D}}}{\longrightarrow} D^m \rightarrow M  \rightarrow 0$ because ${\cal{D}}$ has no CC.  \\   

In addition, it is a fact that has been tested with many students during more than ten years through European international courses, that it is quite difficult to understand certain results that are far from intuition, like the following theorem that can be generalized with the so-called {\it purity filtration} as a way to classify differential modules (See [10], p 201 or [6, 30] for more details):  \\

\noindent
{\bf THEOREM 6.8}: Defining the map $\epsilon$ by $\epsilon (m)(f) = f(m), \forall m \in M, \forall f \in hom_D(M,D) $, we have the exact sequence:  \\
\[   \fbox{  $   0 \longrightarrow t(M) \longrightarrow M \stackrel{\epsilon}{\longrightarrow} {hom}_D({hom}_D(M,D),D)   $  }  \]   
which is explaining why the torsion submodule of $M$ has to do with the kernel of $ad({\cal{D}})$ when $M$ is defined by ${\cal{D}}$, a fact only known in classical control theory when $n = 1$ as we shall see with the example of the double pendulum.  \\  \\

\noindent
{\bf 7) MOTIVATING EXAMPLES}  \\

We present a few examples organized in such a way they end with similar diagrams and open domains for future research as well as test examples for the use of computer algebra.  \\

\noindent
{\bf EXAMPLE 7.1}: ({\it Example 2.1 revisited}) With $m=3, n=1$ and a constant parameter $a$, let us consider the formally surjective  first order operator:   \\
\[  \fbox{  $   {\cal{D}}_1: ( {\eta}^1, {\eta}^2, {\eta}^3 ) \longrightarrow  (d {\eta}^1 - a \, {\eta}^2 - d {\eta}^3= {\zeta}^1, \,\,\, {\eta}^1 - d {\eta}^2 + d {\eta}^3 = {\zeta}^2 )   $  }   \]
Multiplying on the left by two test functions $({\lambda}^1, {\lambda}^2)$ and integrating by parts , we obtain:  \\
\[  \fbox{  $  ad({\cal{D}}_1): ({\lambda}^1, {\lambda}^2) \longrightarrow ( - d {\lambda}^1 + {\lambda}^2 = {\mu}^1, \,\, - a {\lambda}^1 + d {\lambda}^2 = {\mu}^2, \,\, d {\lambda}^1 - d {\lambda}^2 = {\mu}^3 )   $  }  \]
In order to look for the CC of this operator, we obtain first $ - a {\lambda}^1 + {\lambda}^2 = {\mu}^1 + {\mu}^2 + {\mu}^3 \Rightarrow  - a d {\lambda}^1 + d {\lambda}^2 = d{\mu}^1 + 
d {\mu}^2 + d {\mu}^3 \Rightarrow  (a-1) d \lambda \in j_1 (\mu) $. Hence,  $a \neq 1\Rightarrow d \lambda \in j_1 ( \mu)$ and thus $ a \neq 0  \Rightarrow \lambda \in j_1(\mu)$.
Substituting when $a \neq 0, 1$, we obtain therefore the CC operator:   \\
\[   \fbox{  $ ad({\cal{D}}) : ( {\mu}^1, {\mu}^2, {\mu}^3) \longrightarrow  d^2 {\mu}^1 + d^2 {\mu}^2 + d^2 {\mu}^3 - a ( d {\mu}^1 + d {\mu}^2 + {\mu}^3) = \nu  $  }  \]
Multiplying on the left by a test function $\xi$ and integrating by parts, we obtain the second order injective parametrization:  \\
\[  \fbox{  $   {\cal{D}}: \xi  \longrightarrow  ( d^2 \xi + a \, d \xi = {\eta}^1, \,\, d^2 \xi + d \xi = {\eta}^2, \,\, d^2 \xi - a \, \xi = {\eta}^3 )   $  }  \]  \\

\noindent
{\bf EXAMPLE 7.2}: ({\it Bose conjecture}) With $m=n=3$, let $M_1$ be defined by the two PD equations with jet notations $y^3_{12} -y^2_3 - y^3=0, \,\, y^3_{22} - y^1_3=0$. This system is not formally integrable and crossed derivatives provide at once the new second order equation $y^2_{23}- y^1_{13} + y^3_2=0$. The reader could spend hours in order to find out the torsion element $z= y^2_{22}- y^1_{12} + y^1$ that satisfies the autonomous PD equation $z_3=0$. Our purpose is to find it by using only the parametrization test and its corollary. For this, let us multiply on the left the first by the test function ${\lambda}^1$, the second by the test function ${\lambda}^2$, sum and integrate by parts. The adjoint operator is:
\[ \left\{  \begin{array}{lclcl}
y^1 & \longrightarrow & d_3 {\lambda}^2 & = & {\mu}^1  \\
y^2 & \longrightarrow & d_3 {\lambda}^1 & = & {\mu}^2   \\
y^3 & \longrightarrow & d_{22} {\lambda}^2 + d_{12} {\lambda}^1 - {\lambda}^1 & = & {\mu}^3 
\end{array} \right.  \fbox{ $ \begin{array}{ccc}
1 & 2 & 3   \\
1 & 2 & 3 \\
1 & 2 & \bullet
\end{array} $ }   \]
The only CC is $  - d_3 {\mu}^3 + d_{12} {\mu}^2 - {\mu}^2 + d_{22} {\mu}^1 =0 $.  
By double duality, we obtain:  \\
\[  \left\{  \begin{array}{lclcl} 
{\mu}^1 & \longrightarrow & d_{22} \phi & = &  y^1  \\
{\mu}^2 & \longrightarrow & d_{12} \phi - \phi & = &  y^2  \\
{\mu}^3 & \longrightarrow & d_3 \phi & = & y^3
\end{array} \right.   \]
As the left system is not formally integrable, using crossed derivatives on the two first equations, we obtain $ \phi = y^1_{11} -  y^2_{12} - y^2$ and thus {\it at least}  the two CC ([10], Example 5.27, p 219):  \\
\[  \fbox{  $    z \equiv   y^2_{22} - y^1_{12} + y^1 = 0        \hspace{2cm}            y^2_{123} - y^1_{113} + y^2_3 + y^3 =0  $  }  \]
However, such a system is quite far from being even formally integrable and we may not be sure to have {\it generating} CC indeed. Hence, we {\it must} transform the previous system for $\phi$ to an involutive system because no information on generating CC can be known without achieving the PP procedure. The first point is to notice that the parametrization is injective and that the corresponding involutive operator must be $j_2$ in the following diagram using jet notations:  \\ 
\[    \left\{  \begin{array}{ lcl}
{\phi}_{33} & = & y^3_3  \\
{\phi}_{23} & = & y^3_2  \\
{\phi}_{22} & = & y^1  \\
{\phi}_{13} & = & y^3_1  \\
{\phi}_{12} & =  & y^1_{11} - y^2_{12}   \\
{\phi}_{11} & = & y^1_{1111} - y^2_{1112} - y^2_{11}  \\
{\phi}_3 & = &  y^3  \\
{\phi}_2 & = & y^1_1 - y^2_2    \\
{\phi}_1 & =  &  y^1_{111} - y^2_{112} - y^2_1  \\
 \phi & = & y^1_{11} - y^2_{12} - y^2
\end{array} \right.  \fbox{  $ \begin{array}{ccc}
1 & 2 & 3 \\
1 & 2 & \bullet  \\
1 & 2 & \bullet \\
1 & \bullet & \bullet  \\ 
1 & \bullet & \bullet  \\
1 & \bullet & \bullet  \\
\bullet & \bullet & \bullet  \\
\bullet & \bullet & \bullet  \\
\bullet & \bullet & \bullet  \\
\bullet & \bullet & \bullet  
\end{array} $  }    \]
As the generating CC of $j_2$ are produced by the first order operator $D_1$, we can wait for a fourth order system for $y$, {\it at least a third order system}. In fact, {\it though this not evident at at first sight}, the solution we gave in $2001$ of the Bose conjecture implies that the {\it only two} generating CC for the torsion-free differential module $M_1 / t(M_1)$ are the two previous ones. For example, we have:  \\
\[   d_2(y^2_{123} - y^1_{113} + y^2_3 + y^3) - d_{13}(y^2_{22} - y^1_{12} + y^1)= y^2_{23}- y^1_{13} + y^3_2=0   \]
\[  d_2 {\phi}_2 - {\phi}_{22} = d_2 (y^1_1 - y^2_2) - y^1 = - (y^2_{22} - y^1_{12} + y^1) = 0  \]
Such a result is showing the importance of the Spencer operator in actual practice !.  \\

\vspace{2cm}
\noindent
{\bf EXAMPLE  7.3} ({\it RLC electrical circuit})  As we shall prove below, we do believe that the standard control theory of electrical circuits, does not allow {\it at all} to study the structure of the various underlying differential modules defined by the corresponding systems ( torsion submodules, extension modules, resolutions, ... ), in particular if some of the RLC components depend on time.  \\ 
Let us consider as in ([6], p 576) a RLC electrical circuit made up by a battery with voltage $u$ delivering a current $y$ to a parallel subsystem with a branch containing a capacity $C$ with voltage $x^1$ between its two plates and a resistance $R_1$ while the other branch, crossed by a current $x^2$, is containing a coil $L$ and a resistance $R_2$. The three corresponding OD equations are easily seen to be:  \\
\[  \fbox{   $     R_1 C d {x}^1 + x^1 = u, \,\,\,  L d {x}^2 + R_2 x^2 = u, \,\,\,  C d {x}^1 + x^2 = y = - \frac{1}{R_1} x^1 + x^2 + \frac{1}{R_1} u    $   } \]
Such a system can be set up at once in the standard matrix form $\dot{x} =A x + B u, \,\,\, y = Cx + Du $ but we shall avoid the corresponding Kalman criterion that could not be used if $R_1, R_2, L$ or $C$ should depend on time. The two first OD equations are defining a differential module $N$ over the differential field $K= \mathbb{Q}(R_1,R_2,L,C)$ while the elimination of $(x^1,x^2)$ is providing the input submodule $Du=L\subset N$ and the output submodule $Dy=M \subset N$ with $(L,M)\subseteq N$. However, nothing can be said as long as the PP procedure has not been achieved but it has never been used in control theory, in particular for electrical circuit. In the present situation, we have to distinguish carefully between two cases (See [29] for other explicit examples): \\
\noindent
$\bullet$ If \fbox{ $R_1R_2C \neq L$}, we have a single second order CC for $(u,y)$ and the the system is observable, that is we have indeed the strict equality $(L,M)=N$ ({\it Hint}: We let the reader check this fact with $R_1=C=L=1, R_2=2$ and get $ d^2 y + 3 d  y + 2 y - d^2 u - 3 d u - u =0$ which is controllable). \\
\noindent
$\bullet$ If \fbox{ $R_1R_2C=L$}, we have only a single first order equation $L\dot{y} + R_2 y - R_2C \dot{u} - u =0$. Multiplying by a test function $\lambda$ and integrating by parts, we have to solve the two equations $ - L d \lambda -+R_2 |mabda = 0$ and $R_2 C d \lambda - \|ambda = 0$. Hence, this equation is controllable if and only if $L \neq (R_2)^2 C$, thus \fbox{ $R_1\neq R_2$}, and we have the strict inclusion $(L,M)\subset N$ ({\it Hint}: Choose $R_1=R_2=L=C=1$ and get $ \Phi \equiv d^2 y + y - d u -u= 0$ which is not controllable because $z=y-u$ is a torsion element with $ d z+ z=0$). \\
Though it is already quite difficult to find such examples, there is an even more striking fact. Indeed, if we consider only the two first equations for $(x^1, x^2, u)$, we have a formally surjective first order operator ${\cal{D}}$ defined over $K$. Taking into account the intrinsic definition of controllability which is superseding Kalman's one (again because it allows to treat time depending coefficients as well), we let the reader check that the corresponding system is controllable if and only if the first order operator $ad({\cal{D}})$ is injective. Indeed, multiplying the first OD equation by a test function ${\lambda}^1$ and the second by ${\lambda}^2$, we get for the kernel of the adjoint operator: \\
\[ x^1\longrightarrow    - R_1 C d {\lambda}^1 + {\lambda}^1 = 0, \,\, \, x^2 \longrightarrow  - L d {\lambda}^2 + R_2 {\lambda}^2 = 0,\, \,\, 
u \longrightarrow  {\lambda}^1 + {\lambda}^2 = 0 \]
This system is clearly not formally integrable because we obtain by elimination $ L {\lambda}^1 + R_1 R_2 C {\lambda}^2 =0$ and thus $\lambda = 0 $ iff $R_1 R_2 C \neq L$ which is the only condition insuring that $t(N)=0$ a result that must be compared with the Kalman procedure !. \\
Finally, we have to study the differential correspondence between $(x^1, x^2)$ and $(u, y)$, that is to eliminate $(x^1, x^2)$ in order to find the resolvent system  for $ (u,y)$. First of all, we have  $R_1 y = - x^1 + R_1 x^2 + u $ and obtain successively:    
\[  R_1 d y = - d x^1 + R_1 d x^2 + d u \Rightarrow R_1 d y = \frac{1}{R_1 C} x^1 - \frac{R_1 R_2}{L} x^2 + d u - \frac{1}{R_1C} u + \frac{R_1}{L} u  \]
The determinant of this linear system with respect to $(x^1, x^2)$ is vanishing if and only if $R_1 R_1 C = L$. In this case, eliminating $(x^1, x^2)$ by linear combination provides the only single first order OD equation for defining $(L,M)$ by the differential residue $(Dy + Du)/D \Phi$ and $(x^1, x^2)$ cannot be recovered from $(u, y)$. Otherwise, we may exhibit $x^1$ and $x^2$ separately in order to find the second order resolvent system for $(u, y)$ (See [6] for other examples).  \\
As noticed in [5] for the {\it Backlund problem}, we point out the fact that the best way to study a  differential correspondence is to apply the PP procedure to the system in solved form:  \\
\[  d x^2 + \frac{R_2}{L} x^2 = \frac{1}{L} u, \,\,\, d x^1 = \frac{1}{R_1 C} x^1 = \frac{1}{R_1 C}, \,\,\,  x^2  - \frac{1}{R_1} x^1 = y - \frac{1}{R_1} u  \]
These new results could be extended to time dependent electrical components and open a large domain for future control research on electrical circuits.  \\

\vspace{2cm}

\noindent
{\bf EXAMPLE 7.4}: Let $n=2, m= 1$ and introduce the trivial vector bundle $E$ with local coordinates $(x^1, x^2, \xi)$ for a section over the base manifold $X$ with local coordinates $(x^1,x^2)$. Let us consider the linear second order system $R_2 \subset J_2(E)$ defined by the two linearly independent equations $ d_{22}\xi= 0, \,\, d_{12}\xi + a d_1 \xi = 0$ where $a$ is an arbitrary constant parameter. Using crossed derivatives, we get the second order system $R^{(1)}_2 \subset R_2$ defined by the PD equations $d_{22} \xi = 0, d_{12} \xi + a d_1 \xi= 0, a^2 d_1\xi=0$ which is easily seen not to be involutive. Framing essential results and Janet tabulars, we have two possibilities:  \\
 $\bullet  \,\,\, a=0 $: We obtain the following second order homogeneous involutive system: \\
\[ R^{(1)}_2 = R_2 \subset J_2(E) \hspace{2cm} \left\{  \begin{array}{rcl}
 d_{22} \xi &=  & {\eta}^2 \\
d_{12} \xi  & = & {\eta}^1  
\end{array}
\right.  \fbox{ $ \begin{array}{ll}
1 & 2   \\
1 & \bullet 
\end{array} $ } \] 
with the only first order homogeneous involutive CC $d_1 {\eta}^2 - d_2 {\eta}^1= 0 $ leading to the Janet sequence:  \\
\[ \fbox{ $  0 \rightarrow  \Theta \rightarrow E \underset 2 {\stackrel{{\cal{D}}}{\longrightarrow }} F_0   \underset 1 {\stackrel{{\cal{D}}_1}{\longrightarrow }} F_1 
 \rightarrow 0 $ }  \]
We let the reader check as an easy exercise that $ad({\cal{D}})$ which is of order $2$ does not generate the CC of $ad({\cal{D}}_1)$ which is of order $1$. \\

\noindent
$\bullet  \,\,\, a \neq 0$: We obtain the second order system $R^{(1)}_2 $ defined by $ d_{22} \xi = 0, d_{12} \xi = 0, d_1\xi = 0 $ with a strict inclusion $R^{(1)}_2 \subset R_2$ because $3<4$. We may define $\eta = d_1{\eta}^2 - d_2 {\eta}^1+ a {\eta}^1$ and obtain the involutive and finite type system in $\delta$-regular coordinates:  \\
\noindent
\[ R^{(2)}_2 \subset J_2(E) \hspace{2cm} \left\{  \begin{array}{rcl}
 d_{22} \xi &=  & {\eta}^2  \\
 d_{12} \xi & =  & {\eta}^1 - \frac{1}{a} \eta   \\
 d_{11} \xi & =  & \frac{1}{a^2} d_1 \eta \\
 d_1  \xi    & = & \frac{1}{a^2} \eta
\end{array}
\right.  \fbox{ $ \begin{array}{ll}
1 & 2   \\
1 & \bullet \\  
1 & \bullet  \\
\bullet & \bullet 
\end{array} $ } \]
Counting the dimensions, we have the following strict inclusions by comparing the dimensions:  \\
\[      R^{(2)}_2 \subset R^{1)}_2 \subset R_2 \subset J_2(E) , \hspace{3cm}  2 <  3   <  4  <  6  \]
We have proved  symbol $g_{2+r} $ is involutive with $dim(g_{2+r})=1, \forall r \geq 0$ and that $dim(R_{2+r})  = 4, \forall r\geq 0$. \\
After differentiating twice, we could be waiting for CC of order $3$. However, we obtain the $4$ CC:  \\
\[ d_2 {\eta}^1 - \frac{1}{a} d_2 \eta - d_1 {\eta}^2 = 0, \frac{1}{a^2} d_{12} \eta - d_1 {\eta}^1 + \frac{1}{a} d_1\eta = 0, \frac{1}{a^2} d_2 \eta - {\eta}^1 + \frac{1}{a} \eta =0, 
\frac{1}{a^2} (d_1 \eta - d_1 \eta) = 0  \]
The last CC that we shall call "{\it identity to zero} " must not be taking into account. The second CC is just the derivative with respect to $x^1 $ of the third CC which amounts to 
\[ (d_{12} {\eta}^2 - d_{22}{\eta}^1 + a d_2 {\eta}^1) - a^2 {\eta}^1 + a (d_1 {\eta}^2 - d_2 {\eta}^1 + a {\eta}^1)=0 \Leftrightarrow  d_{12} {\eta}^2 - d_{22} {\eta}^1 + a d_1 {\eta}^2=0 \]
which is a second order CC amounting to the first. Hence we get the only formally surjective generating CC operator: 
\[  \fbox{  $   {\cal{D}}_1: ({\eta}^1, {\eta}^2) \rightarrow d_{12} {\eta}^2 - d_{22} {\eta}^1 + a d_1 {\eta}^2=\zeta  $  }   \]
 As a byproduct we have the exact sequences 
$\forall r\geq 0$:  \\
\[   0 \rightarrow  R_{r + 4} \rightarrow J_{r + 4}(E) \rightarrow  J_{r + 2}(F_0) \rightarrow  J_r(F_1) \rightarrow 0   \] 
Such a result can be checked directly through the identity:   \\
 \[   4 - (r+5)(r+6)/2 + 2(r+3)(r+4)/2 - (r+1)(r+2)/2 = 0   \]
We obtain therefore the formally exact sequence we were looking for, namely:  \\
\[ \fbox{ $ 0 \rightarrow \Theta \rightarrow E \underset 2{\stackrel{{\cal{D}}}{\longrightarrow}}  F_0  \underset 2{\stackrel{{\cal{D}}_1}{\longrightarrow}}  F_1 
\rightarrow 0 $ }  \]
The {\it surprising fact} is that, {\it in this case}, $ad({\cal{D}}) $ generates the CC of $ad({\cal{D}}_1) $. Indeed, multiplying by the Lagrange multiplier test function $\lambda$ and integrating by parts, we obtain the second order operator:  
\[  \fbox{  $  ad({\cal{D}}_1): \lambda \longrightarrow (- d_{22} \lambda = {\mu}^1, d_{12} \lambda - a d_1 \lambda = {\mu}^2 ) $  }   \]
and thus $ - a^2 d_1 \lambda = d_1 {\mu}^1 + d_2 {\mu}^2  + a {\mu}^2 $. Substituting, we finally get the only second order CC  operator: 
\[  \fbox{  $   ad({\cal{D}}):  ({\mu}^1, {\mu}^2) \longrightarrow   d_{12} {\mu}^1 + d_{22}{\mu}^2 - a d_1 {\mu}^1=\nu   $   }  \]                                                                                                
We have the long exact sequence and its adjoint sequence which is also exact:
\[   \fbox{  $         \begin{array}{rcccccccl}
  & & 1 & \underset 2{\stackrel{{\cal{D}}}{\longrightarrow}} & 2 & \underset 2{\stackrel{{\cal{D}}_1}{\longrightarrow} }  &  1  & \longrightarrow   & 0  \\
&  &  &  &  &  &  &  &    \\
 0  & \longleftarrow   & 1 & \underset 2{\stackrel{ad({\cal{D}})}{\longleftarrow}} & 2 & \underset 2{\stackrel{ad({\cal{D}}_1)}{\longleftarrow} }  &  1  &  &    
 \end{array}     $  }       \]  
  \hspace{2cm}

Contrary to what happens in OD control theory, in the present situation with $n=2$, ${\cal{D}}_1$ may be formally surjective and may admit a parametrization but its adjoint 
may not be injective.  \\   \\

\noindent
{\bf EXAMPLE 7.5}: The following example illustrates the concept of parametrization, showing in particular that different parametrizations may exist that may not be minimum parametrizations.  \\

 Let us consider the first order operator with two independent variables $(x^1,x^2)$:   \\ 
\[ \fbox{  $ {\cal{D}}_1: ({\eta}^1, {\eta}^2) \rightarrow d_2{\eta}^1 - d_1{\eta}^2 + x^2 {\eta}^2=\zeta  $  }   \]
The ring of differential operators involved is $D=K[d_1,d_2]$ with $K=\mathbb{Q}(x^1,x^2)= \mathbb{Q} (x)$. Multiplying on the left by a test function $\lambda$ and integrating by parts, we get the adjoint operator:  \\
\[ \fbox{  $   ad({\cal{D}}_1): \lambda \rightarrow ( - d_2 \lambda = {\mu}^1, d_1\lambda + x^2  \lambda = {\mu}^2 )   $  }  \]
Using crossed derivatives, this operator is injective because $\lambda= d_2{\mu}^2+ d_1{\mu}^1+x^2{\mu}^1$ and we even obtain a lift  $\lambda \longrightarrow \mu \longrightarrow \lambda$. Substituting, we get $ad({\cal{D}}): ({\mu}^1,{\mu}^2) \rightarrow ({\nu}^1,{\nu}^2)$:  \\
\[  \fbox{  $   d_{22}{\mu}^2 + d_{12}{\mu}^1+x^2 d_2{\mu}^1 + 2{\mu}^1  =  {\nu}^1, d_{12}{\mu}^2 + d_{11}{\mu}^1+2x^2d_1{\mu}^1+x^2 d_2{\mu}^2+(x^2)^2{\mu}^1-{\mu}^2 =  {\nu}^2  $  }  \] 
allowing to define a second order operator ${\cal{D}}$ by using the fact that $ad(ad({\cal{D}}))= {\cal{D}}$. This operator is involutive and the only corresponding generating CC is $ d_2{\nu}^2 - d_1{\nu}^1 - x^2 {\nu}^1=0 $.
Therefore ${\nu}^2$ is differentially dependent on ${\nu}^1$ but ${\nu}^1$ is also differentially dependent on ${\nu}^2$. Multiplying on the left by a test function $\theta$ and integrating by parts, the corresponding adjoint operator is:  \\
\[  \fbox{  $  {\cal{D}}_{-1}: \theta \rightarrow ( d_1 \theta - x^2 \theta ={\xi}^1,  - d_2 \theta    ={\xi}^2 )  $  } \]
Multiplying now the first equation of $ad({\cal{D}})$ by the test function ${\xi}^1$, the second equation by the test function ${\xi}^2$, adding and integrating by parts, we get the second order operator :  \\
\[ \fbox{  $  {\cal{D}}: ({\xi}^1, {\xi}^2) \rightarrow  \left\{ \begin{array}{lcl}
 d_{22} {\xi}^1 + d_{12} {\xi}^2 - x^2 d_2 {\xi}^2 - 2 {\xi}^2 & = & {\eta}^2  \\
  d_{12}{\xi}^1 + d_{11}{\xi}^2 - x^2 d_2{\xi}^1 - 2x^2 d_1{\xi}^2+{\xi}^1 +(x^2)^2{\xi}^2  & = & {\eta}^1 
 \end{array} \right. $  }     \]      
which is easily seen to be a parametrization of ${\cal{D}}_1$. This operator is involutive and the kernel of this parametrization has differential rank equal to $1$ because ${\xi}^1$ or ${\xi}^2$ can be given arbitrarily.  \\
We can now consider each component of $\xi$ {\it separately}. Keeping for example ${\xi}^1=\xi$ while setting ${\xi}^2=0$, we get the {\it first second order minimal parametrization} 
\[ \fbox{  $  \xi \rightarrow ( d_{22}\xi = {\eta}^2 , \, \, \, d_{12}\xi - x^2 d_2\xi + \xi ={\eta}^1)  $  }   \]
This system is again involutive and the parametrization is minimal because the kernel of this parametrization has differential rank equal to $0$. With a similar comment, setting now ${\xi}^1=0$ while keeping ${\xi}^2={\xi}'$, we get the {\it second second order minimal parametrization}:  \\
\[   \fbox{  $    {\xi}' \rightarrow ( d_{11}{\xi}'-2x^2 d_1{\xi}'+(x^2)^2{\xi}' = {\eta}^1 , \, \, \,  d_{12}{\xi}'  - x^2 d_2{\xi}' - 2{\xi}' = {\eta}^2 )   $   }  \]
which is again easily seen to be involutive by exchanging $x^1$ with $x^2$.  \\
With again a similar comment, setting now ${\xi}^1= d_1\phi, {\xi}_2= - d_2\phi$ in the canonical parametrization, we obtain the {\it third different second order minimal parametrization}:  \\
\[  \fbox{  $ \phi \rightarrow  ( x^2 d_{22}\phi + 2 d_2\phi = {\eta}^2, \, \, \, x^2 d_{12}\phi - (x^2)^2 d_2\phi + 
d_1\phi = {\eta}^1)  $  }   \]

We have the long exact sequence and its adjoint sequence which is also exact:
\[   \fbox{  $         \begin{array}{rccccccccl}
0  & \longrightarrow &1& \underset 1{ \stackrel{{\cal{D}}_{-1} }{\longrightarrow} }  & 2 & \underset 2{\stackrel{{\cal{D}}}{\longrightarrow}} & 2 & \underset 1{\stackrel{{\cal{D}}_1}{\longrightarrow} }  &  1  & \longrightarrow    0  \\
&  &  &  &  &  &  &  &  &   \\
 0  & \longleftarrow &1& \underset 1{ \stackrel{ad({\cal{D}}_{-1} )}{\longleftarrow} }  & 2 & \underset 2{\stackrel{ad({\cal{D}})}{\longleftarrow}} & 2 & \underset 1{\stackrel{ad({\cal{D}}_1)}{\longleftarrow} }  &  1  & \longleftarrow    0   
 \end{array}     $  }       \]  
We notice that ${\cal{D}}_1$ is parametrized by ${\cal{D}}$ which is again parametrized by ${\cal{D}}_{-1}$, exactly like $div$ is parametrized by $curl$ which is again parametrized by $grad$ in classical vector geometry. In the present example, one can prove that there is an isomorphism $ M_1 \simeq hom_D(hom_D(M_1,D),D)$ and $M_1$ is called a "{\it reflexive}" differential module (See [10], p 200 for more details).     \\   

\vspace{2cm} 

\noindent
{\bf EXAMPLE 7.6}: ({\it Contact transformations}): With $m=n=3, q=1, K=\mathbb{Q}(x^1,x^2,x^3) = \mathbb{Q}(x)$, we may introduce the $1$-form $\alpha=dx^1-x^3dx^2 \in T^*$ and consider the Lie pseudogroup of transformations preserving $\alpha$ up to a function factor, defined by $j_1(f)^{-1}(\alpha)=\rho(x) \alpha$ that is to say $ {\alpha}_k(f(x)) {\partial}_i f^k(x) = \rho (x) {\alpha}_i(x)$. Also, $\alpha \wedge d \alpha = d x^1 \wedge d x^2 \wedge d x^3 = d x \Rightarrow j_1(f)^{-1} (d x) = {\rho}^2(x) d x$. Eliminating the factor $\rho (x)$ and linearizing at  the identity, we obtain a first order system of infinitesimal Lie equations which is not even formally integrable and must use the PP procedure to get an involutive system with corresponding Janet tabular:  \\
\[ \fbox{  $  {\cal{D}}  \hspace{2cm}  \left\{ \begin{array}{lcr}
{\partial}_3{\xi}^3 + {\partial}_2 {\xi}^2 - {\partial}_1{\xi}^1+ 2 x^3 {\partial}_1{\xi}^2 & = &  {\eta}^3  \\
{\partial}_3{\xi}^1 - x^3 {\partial}_3{\xi}^2  & =   & {\eta}^2 \\
 {\partial}_2{\xi}^1 - x^3 {\partial}_2{\xi}^2 +x^3 {\partial}_1{\xi}^1 - (x^3)^2{\partial}_1{\xi}^2 - {\xi}^3 & =  & {\eta}^1
\end{array}  \right.  \fbox{$ \begin{array}{lll}
1 & 2 & 3  \\
1 & 2 & 3  \\
1 & 2 & \bullet
\end{array} $ }  $  }  \]
There is thus one CC of order $1$ described by the formally surjective first order operator ${\cal{D}}_1$: 
\[    \fbox{  $  {\cal{D}}_1  \hspace{2cm}   d_3{\eta}^1-d_2{\eta}^2 - x^3 d_1{\eta}^2 + {\eta}^3 = \zeta  $  }  \]
Multiplying $ ({\eta}^1, {\eta}^2, {\eta}^3)$ by the test functions $({\mu}^1, {\mu}^2, {\mu}^3)$ and integrating by parts, we obtain ({\it by chance} !) the involutive operator:   \\
\[   \fbox{  $  ad({\cal{D}})   \hspace{2cm}\left\{ \begin{array}{lcl}
{\partial}_3{\mu}^3 + {\mu}^1 & = & - {\nu}^3  \\
{\partial}_3{\mu}^2 + x^3 {\partial}_1 {\mu}^1 + {\partial}_2 {\mu}^1 - {\partial}_1 {\mu}^3  & =   & - {\nu}^1 \\
 {\partial}_2{\mu}^3 + x^3 {\partial}_1 {\mu}^3 - {\mu}^2 & =  & - ({\nu}^2 + x^3 {\nu}^1)
\end{array}  \right.  \fbox{$ \begin{array}{lll}
1 & 2 & 3  \\
1 & 2 & 3  \\
1 & 2 & \bullet
\end{array} $ } $  } \]
providing the only first order CC:  \\
\[ \fbox{  $  ad({\cal{D}}_{-1}): ({\nu}^1, {\nu}^2, {\nu}^3) \rightarrow  x^3 {\partial}_3 {\nu}^1 + 2 {\nu}^1  + {\partial}_3 {\nu}^2 - {\partial}_2 {\nu}^3 - x^3 {\partial}_1 { \nu}^3 = 0  $  } \]
and the classical injective parametrization:   \\
\[  \fbox{  $  {\cal{D}}_{-1}  \hspace{1cm}  - x^3 {\partial}_3 \phi  + \phi = {\xi}^1, \,\, - {\partial}_3 \phi = {\xi}^2, \,\, {\partial}_2\phi + x^3 {\partial}_1\phi={\xi}^3  \,\, \Rightarrow  \,\, 
{\xi}^1 - x^3 {\xi}^2=\phi   $   } \]

Using the Vessiot structure equations [31], we notice that $\alpha$ is not invariant by the contact Lie pseudogroup. The associated invariant geometric object is a $1$-form density 
$\omega$ leading to the system of infinitesimal Lie equations in Medolaghi form:  \\
\[    \fbox{  $     {\Omega}_i\equiv ( {\cal{L}}(\xi)\omega)_i \equiv {\omega}_r{\partial}_i{\xi}^r - \frac{1}{2}{\omega}_i{\partial}_r{\xi}^r + {\xi}^r{\partial}_r{\omega}_i=0 $  }\]
which becomes formally integrable {\it if and only if} the following only Vessiot structure equation, still not known today, is satisfied:  \\
\[   \fbox{  $    {\omega}_1({\partial}_2{\omega}_3- {\partial}_3{\omega}_2) +   {\omega}_2({\partial}_3{\omega}_1- {\partial}_1{\omega}_3) +  {\omega}_3({\partial}_1{\omega}_2- {\partial}_2{\omega}_3)=c  $  } \]
with the {\it only} structure constant $c$. In the present contact situation, we may choose $\omega=(1, -x^3,0)$ as we did and get $c=1$ but we may also choose $\omega=(1,0,0)$ and get $c=0$. This new choices is also bringing an involutive system:    \
\[  \fbox{ $ -2{\Omega}_1\equiv  {\partial}_3{\xi}^3 + {\partial}_2{\xi}^2 - {\partial}_1{\xi}^1=0,\,\, {\Omega}_2\equiv {\partial}_2{\xi}^1=0, \,\, {\Omega}_3\equiv {\partial}_3{\xi}^1=0  $ }  \]
having the only CC $ d_2{\Omega}_3 - d_3 {\Omega}_2=0 $. However, ${\xi}^1$ is indeed a torsion element and one cannot find a parametrization. Such an example is thus proving that the existence of a parametrization for systems of Lie equations highly depends on the Vessiot structure constants.  \\  \\

\noindent
{\bf EXAMPLE 7.7}: ({\it Double pendulum}) Many examples can be found in classical ordinary differential control theory because it is known that a linear control system is controllable if and only if it is parametrizable (See [6, 10] for more details and examples). In our opinion, the best and simplest one is the so-called double pendulum in which a rigid bar is able to move horizontally with reference position $x$ and we attach two pendulums with respective length $l_1$ and $l_2$ making the (small) angles ${\theta}_1$ and ${\theta}_2$ with the vertical, the corresponding control system does not depend on the mass of each pendulum and the two equations easily follow by projection from the Newton laws:  \\
\[  \fbox{  $ {\cal{D}}_1 \eta = 0 \,\, \Leftrightarrow \,\,\,  d^2x +l_1d^2{\theta}^1 +g {\theta}^1=0, \hspace{1cm}  d^2x + l_2 d^2{\theta}^2 + g{\theta}^2=0  $  }   \]
where $g$ is the gravity. A first result, still not acknowledged by the control community, is to prove that this control system is controllable if and only if $l_1 \neq  l_2$ without using a tedious computation through the standard Kalman test but, {\it equivalently}, to prove that the corresponding second order operator $ad({\cal{D}}_1)$ is injective. Though this is not evident, such a result comes from the fact $D$ is a principal ideal ring when $n = 1$ and thus, if the differential module $M_1$ is torsion-free, then $M_1$ is also free and has a basis allowing to split the short exact resolution $0 \rightarrow D^2 \stackrel{{\cal{D}}_1}{\longrightarrow} D^3 \rightarrow M_1 \rightarrow 0$ with $M_1 \simeq D$ in this case (See [10] p 204-205 or [6] for details).
Hence, multiplying on the left the first OD equation by ${\lambda}^1$, the second by ${\lambda}^2$, then adding and integrating by parts, we get:
 \[  \fbox{  $  ad({\cal{D}}_1) \lambda= \mu  \,\,\, \Leftrightarrow  \,\,\, 
  \left\{  \begin{array}{lcccl}
 x & \longrightarrow  & d^2 {\lambda}^1 + d^2 {\lambda}^2 & = &    {\mu}^1  \\
 {\theta}^1 & \longrightarrow &  l_1 d^2 {\lambda}^1   + g {\lambda}^1 & = & {\mu}^2 \\
 {\theta}^2  & \longrightarrow  &  l_2 d^2 {\lambda}^2  + g {\lambda}^2 &  =  & {\mu}^3
\end{array}  \right.  $   }    \]
The main problem is that the operator $ad({\cal{D}}_1)$ is {\it not} formally integrable because we have: 
\[   g(l_2 {\lambda}^1 + l_1 {\lambda}^2) = l_2 {\mu}^2 + l_1 {\mu}^3 - l_1 l_2 {\mu}^1  \]
and is thus injective if and only if $l_1 \neq l_2$ because, differentiating twice this equation, we also get:
\[       (l_2 / l_1) {\lambda}^1 + (l_1 / l_2) {\lambda}^2   \in j_2 (\mu)      \]
Hence, if $l_1 \neq l_2$, we finally obtain $ \lambda \in j_2 (\mu)$ and, after substitution, a single fourth order CC for $\mu$ showing that $ad({\cal{D}})$ is indeed a fourth order operator, {\it a result not evident indeed at first sight}. It follows that we have thus been able to work out the parametrizing operator ${\cal{D}}$ of order 4, namely:   \\
\[  \fbox{  $   {\cal{D}} \phi=  \eta \,\,\,  \Leftrightarrow  \,\,\,    \left\{  \begin{array}{rcl}
  - l_1 l_2 d^4\phi - g(l_1+l_2)d^2 \phi - g^2 \phi & = & x  \\
  l_2 d^4\phi +g d^2 \phi & = & {\theta}_1  \\
  l_1 d^4 \phi + gd^2 \phi  & = & {\theta}_2
  \end{array} \right.  $   }   \]  
 This parametrization is injective iff $l_1 \neq l_2$ because we have successively with $g\neq 0$:  \\
 \[     l_2 d^2 \phi + g \phi = 0, \Rightarrow   l_1 d^2 \phi +g \phi = 0 \Rightarrow    g(l_1 - l_2) \phi =0 \Rightarrow  \phi = 0  \]  
We have the long exact sequence and its adjoint sequence which is also exact:
\[   \fbox{  $         \begin{array}{rcccccccl}
0  & \longrightarrow  & 1 & \underset 4{\stackrel{{\cal{D}}}{\longrightarrow}} & 3 & \underset 2{\stackrel{{\cal{D}}_1}{\longrightarrow} }  &  2  & \longrightarrow   &  0  \\
&  &  &  &  &  &  &  &    \\
 0  & \longleftarrow   & 1 & \underset 4{\stackrel{ad({\cal{D}})}{\longleftarrow}} & 3 & \underset 2{\stackrel{ad({\cal{D}}_1)}{\longleftarrow} }  &  2  & \longleftarrow  &  0   
 \end{array}     $  }       \]

  We finally consider the case $l_1= l_2 = l$. Substracting the two OD equations, we discover that $ z = {\theta}^1 - {\theta}^2$ is an observable quantity that satisfies the autonomous system $ l d^2 z + g z = 0$ existing for a single pendulum. It follows that $ z $ is a torsion element and the system cannot be controllable. When $ z = 0 \Rightarrow {\theta}^1 = {\theta}^2 = \theta$ we let the reader prove that the remaining OD equation $d^2 x + l d^2 {\theta} + g {\theta} = 0$ can be parametrized by $l d^2 \xi + g \xi = x, - d^2 \xi = {\theta}$. \\
  
 Comparing this approach to the standard Kalman procedure that can be found in all textbooks, such an example is proving that {\it the mathematical foundations of control theory must be entirely revisited because controllability is a built in property of a control system, not depending on the choice of inputs and outputs among the system variables}.  \\

At this stage of the reading, we invite the reader to realize this experiment with a few dollars, check how the controllability depends on the lengths and wonder how this example may have ANYTHING to do with the Cosserat, Einstein or Maxwell equations !.  \\

 \noindent
  {\bf EXAMPLE 7.8}: ({\it Einstein equations}) A less academic and more difficult example is proving that {\it Einstein equations cannot be parametrized contrary to Maxwell equations}. The {\it Einstein} operator is self-adjoint, that is $ad(Einstein) = Einstein$ (a crucial property indeed, for which you will not find any reference !!!), we obtain successively [10, 32]: \\
  \[  \fbox{1}    {\cal{D}}_1 = Einstein, \fbox{2}  ad({\cal{D}}_1) = Einstein, \fbox{3}  ad({\cal{D}}) = Cauchy, \fbox{4}  {\cal{D}} = Killing, \fbox{5}  {\cal{ D}}_1'  = Riemann    \]
 and we obtain thus the strict symbolic inclusion ${\cal{D}}_1 \subset  {\cal{D}}'_1$ by counting the number of CC as $n(n+1)/2 < n^2 (n^2 - 1)/12, \forall n \geq 4$ along the following diagram existing when $n=4$:  \\  
  
\[ \fbox{  $ \begin{array}{rcccccccl}
  & &  &  &  &  &  &  &  \\
 & &  & 10 &\stackrel{Riemann}{\longrightarrow}  & 20 & \stackrel{Bianchi}{\longrightarrow}  & 10 & \\
 & & & \parallel & \nearrow & \downarrow &  & \downarrow &  \\
 & 4 &  \stackrel{Killing}{\longrightarrow} & 10 & \stackrel{Einstein}{\longrightarrow} & 10  & \stackrel{div}{\longrightarrow} &  4  & \longrightarrow 0 \\
  & & &  & &  &  & &  \\
 0 \longrightarrow & 4 & \stackrel{Cauchy}{\longleftarrow} & 10 & \stackrel{Einstein}{\longleftarrow} & 10 &   & &  \\
  & & & & & & & & 
\end{array} $  }   \]  \\
It follows that the {\it Cauchy} and {\it Killing} operators ({\it left side}) has {\it strictly nothing to do} with the {\it Bianchi} and thus {\it div} operators ({\it right side}). In addition, the $10$ stress potentials are no longer tensors but tensor densities and have nothing to do with the perturbation of the metric. According to the last corollary, the $20 - 10 = 10$ new CC are generating the torsion submodule of the differential module defined by the Einstein operator. One can prove that such a basis is made by the $10$ independent components of the Weyl tensor, each one being killed by the second order Dalembertian, a result leading to the so-called {\it Lichnerowicz waves} (in France !) [33] and {\it totally unknown} in this differential module framework !!!. The specific cases $n= 2 $ and $n= 3$ will be considered later on.  \\  \\

\noindent
{\bf EXAMPLE 7.9} ({\it Maxwell equations}) When $n=4$, a similar comment can be done for electromagnetism through the exterior derivative as the first set of Maxwell equations 
$  d: {\wedge}^2 T^*  \rightarrow {\wedge}^3 T^* $ can be parametrized by the EM potential 1-form with $ d: T^* \rightarrow {\wedge}^2 T^* : A  \rightarrow d A = F$, while the second set of Maxwell equations (adjoint of this parametrization) $ ad(d)= {\wedge}^4 T^* \otimes {\wedge}^2 T \rightarrow   {\wedge}^4 \otimes T : {\cal{F}} \longrightarrow  {\cal{J}}$ can be parametrized by the EM pseudo-potential $ ad(d): {\wedge}^4 T^* \otimes  {\wedge}^3 T \longrightarrow {\wedge}^4 T^* \otimes {\wedge}^2 T$ and we have the exact Poincar\'{e} sequence and its adjoint  sequence  which is also exact:
\[   \fbox{   $  \begin{array}{ccccc}
  A  &  \longrightarrow & F &  &  \\
  4 & \underset 1{\stackrel{d}{\longrightarrow}} & 6 & \underset 1{\stackrel{d}{\longrightarrow}} & 4   \\
   & & &  &      \\
 4 & \underset 1{\stackrel{ad(d)}{\longleftarrow}} &  6   &  \underset 1{\stackrel{ad(d) }{\longleftarrow}} &  4  \\
 {\cal{J}} & \longleftarrow  &  {\cal{F}}  &  &
\end{array}  $  }  \]

These results are even strengthening the comments we shall make in the next section on the origin of gravitational waves [27, 28].  \\

\noindent
{\bf 8) RIEMANNIAN GEOMETRY REVISITED} \\

Linearizing the {\it Ricci} tensor ${\rho}_{ij}$ over the Minkowski metric $\omega$, we obtain the usual second order homogeneous {\it Ricci} operator $\Omega \rightarrow R$ with $4$ terms, {\it which is not self-adjoint} [32]:  \\
\[  2 R_{ij}= {\omega}^{rs}(d_{rs}{\Omega}_{ij}+ d_{ij}{\Omega}_{rs} - d_{ri}{\Omega}_{sj} - d_{sj}{\Omega}_{ri})= 2R_{ji}  \]
\[ tr(R)= {\omega}^{ij}R_{ij}={\omega}^{ij}d_{ij}tr(\Omega)-{\omega}^{ru}{\omega}^{sv}d_{rs}{\Omega}_{uv}  \]
We may define the $Einstein$ operator by setting $E_{ij}=R_{ij} - \frac{1}{2} {\omega}_{ij}tr(R)$ and obtain the $6$ terms, {\it which, surprisingly, is self-adjoint} [GR]:  \\
\[ 2E_{ij}= {\omega}^{rs}(d_{rs}{\Omega}_{ij} + d_{ij}{\Omega}_{rs} - d_{ri}{\Omega}_{sj} - d_{sj}{\Omega}_{ri})
- {\omega}_{ij}({\omega}^{rs}{\omega}^{uv}d_{rs}{\Omega}_{uv} - {\omega}^{ru}{\omega}^{sv}d_{rs}{\Omega}_{uv})  \]
We have the (locally exact) differential sequence of operators acting on sections of vector bundles where the order of an operator is written under its arrow.:  \\
\[    T \underset 1{\stackrel{Killing}{\longrightarrow}} S_2T^* \underset 2{\stackrel{Riemann}{\longrightarrow}} F_1 \underset 1{\stackrel{Bianchi}{\longrightarrow}} F_2  \]
\[    n \stackrel{{\cal{D}}}{ \longrightarrow} n(n+1)/2 \stackrel{{\cal{D}}_1}{\longrightarrow} n^2(n^2-1)/12 \stackrel{{\cal{D}}_2}{\longrightarrow} n^2(n^2-1)(n-2)/24  \]
Our purpose is now to study the differential sequence onto which its right part is projecting:  \\
\[      S_2T^* \underset 2 {\stackrel{Einstein}{\longrightarrow}} S_2T^* \underset 1{\stackrel{div}{\longrightarrow}} T^*  \rightarrow 0  \]
\[  n(n+1)/2 \longrightarrow  n(n+1)/2  \longrightarrow n  \rightarrow 0  \]
and the following adjoint sequence where we have set [LANC]] :    \\
\[   ad(T) \stackrel{Cauchy}{\longleftarrow} ad(S_2T^*)  \stackrel{Beltrami}{\longleftarrow} ad(F_1) \stackrel{Lanczos}{\longleftarrow} ad(F_2)     \]
In this sequence, if $E$ is a vector bundle over the ground manifold $X$ with dimension $n$, we may introduce the new vector bundle $ad(E)={\wedge}^nT^* \otimes E^*$ where $E^*$ is obtained from $E$ by inverting the transition rules exactly like $T^*$ is obtained from $T$. We have for example $ad(T)={\wedge}^nT^*\otimes T^*\simeq {\wedge}^nT^*\otimes T \simeq {\wedge}^{n-1}T^*$ because $T^*$ is isomorphic to $T$ by using the metric $\omega$. The $Einstein$ operator is induced from the $Riemann$ operator and the $div$ operator matrix is induced from the $Bianchi$ operator by contracting indices. We advise the reader not familiar with the formal theory of systems or operators to follow the computation in dimension $n=2$ with the  $Airy$ operator, which is the formal adjoint of the $Riemann$ operator, and $n=3$ with the $Beltrami$ operator which is the formal adjoint of the $Riemann$ operator which will be seen to be self-adjoint up to a change of basis. With more details, we have:  \\

\noindent
$\bullet \hspace{3mm}n=2$: The stress equations become $ d_1{\sigma}^{11}+ d_2{\sigma}^{12}=0, d_1{\sigma}^{21}+ d_2{\sigma}^{22}=0$. Their second order parametrization ${\sigma}^{11}= d_{22}\phi, {\sigma}^{12}={\sigma}^{21}= - d_{12}\phi, {\sigma}^{22}= d_{11}\phi$ has been provided by George Biddell Airy in 1863 [34] and is well known in plane elasticity, for example when constructing a dam (See the Introduction of [6]). We get the second order system with Janet tabular:  \\
\[ \left\{  \begin{array}{rll}
{\sigma}^{11} & \equiv d_{22}\phi =0 \\
-{\sigma}^{12} & \equiv d_{12}\phi =0 \\
{\sigma}^{22} & \equiv d_{11}\phi=0
\end{array}
\right. \fbox{ $ \begin{array}{ll}
1 & 2   \\
1 & \bullet \\  
1 & \bullet  
\end{array} $ } \]
which is involutive with one equation of class $2$, $2$ equations of class $1$ and the $2$ corresponding first order CC are just the $Cauchy$  equations. Of course, the $Airy$ function ($1$ term) has absolutely nothing to do with the perturbation of the metric ($3$ terms). Indeed, when $\omega$ is the Euclidean metric, we may consider the only component:   \\
\[  \begin{array}{rcl}
tr(R)  &  = &  (d_{11} + d_{22})({\Omega}_{11} + {\Omega}_{22}) - (d_{11}{\Omega}_{11} + 2 d_{12}{\Omega}_{12}+ d_{22}{\Omega}_{22})  \\
      &  =  &  d_{22}{\Omega}_{11} + d_{11}{\Omega}_{22} - 2 d_{12}{\Omega}_{12}
      \end{array}  \]
Multiplying by the Airy function $\phi$ and integrating by parts, we discover that:   \\
      \[    Airy=ad(Riemann) \,\,\,\,  \Leftrightarrow  \,\,\,\,  Riemann = ad(Airy)   \]  \\
in the following exact sequence and its exact adjoint sequence:
 \[ \fbox{  $ \begin{array}{rcccccccl}
    &  &  2  &  \underset 1 {\stackrel{Killing}{\longrightarrow}} & 3 &  \underset 2 {\stackrel{Riemann}{\longrightarrow}} &1 & \longrightarrow  &   0  \\
     &  &  &  &  &  &  &  &  \\
  0  & \longleftarrow   & 2 & \underset 1 {\stackrel{Cauchy}{\longleftarrow}}  &3  & \underset 2 {\stackrel{Airy}{\longleftarrow}}  & 1 & &   
      \end{array}  $  }  \]
      
\noindent
$\bullet \hspace{3mm} n=3$: It is quite more delicate to parametrize the $3$ PD equations: \\
\[ d_1{\sigma}^{11}+ d_2{\sigma}^{12}+ d_3{\sigma}^{13}=0,\hspace{3mm} d_1{\sigma}^{21}+ d_2{\sigma}^{22}+ d_3{\sigma}^{23}=0, \hspace{3mm} d_1{\sigma}^{31}+ d_2{\sigma}^{32}+ d_3{\sigma}^{33}=0 \]
A direct computational approach has been provided by Eugenio Beltrami in 1892 [35], James Clerk Maxwell in 1870 [36] and Giacinto Morera in 1892 [37] by introducing the $6$ {\it stress functions} ${\phi}_{ij}={\phi}_{ji}$ in the {\it Beltrami parametrization}:    \\
\[  \fbox{  $  \left\{  \begin{array}{rll}
{\sigma}^{11} \equiv & d_{33}{\phi}_{22}+ d_{22}{\phi}_{33}-2 d_{23}{\phi}_{23} = 0  \\
{\sigma}^{12}\equiv & d_{13}{\phi}_{23}+ d_{23}{\phi}_{13} - d_{33}{\phi}_{12}- d_{12}{\phi}_{33} =0  \\
 {\sigma}^{13}\equiv & d_{23}{\phi}_{12}+ d_{12}{\phi}_{23} - d_{22}{\phi}_{13}- d_{13}{\phi}_{22} =0 \\
 {\sigma}^{22}\equiv & d_{33}{\phi}_{11}+ d_{11}{\phi}_{33} -2 d_{13}{\phi}_{13} = 0  \\
{\sigma}^{23}\equiv & d_{12}{\phi}_{13}+ d_{13}{\phi}_{12} - d_{23}{\phi}_{11}- d_{11}{\phi}_{23} = 0 \\
{\sigma}^{33}\equiv & d_{22}{\phi}_{11}+ d_{11}{\phi}_{22}-2 d_{12}{\phi}_{12} = 0
\end{array} \right.  $    }    \]  \\
Changing the sign of the second and fifth equations, this system is involutive with $3$ equations of class $3$, $3$ equations of class $2$ and no equation of class $1$. The corresponding Janet tabular is:  
\[  \fbox{  $   \begin{array}{ccc}
 1 & 2 & 3    \\
 1 & 2 & 3    \\
  1 & 2 & 3  \\
  1 & 2 & \bullet  \\
  1 & 2 & \bullet  \\
  1 & 2 & \bullet  
\end{array}  $  }  \]

The $3$ CC are describing the stress equations which admit therefore a parametrization ... but without any geometric framework, in particular without any possibility to imagine that the above second order operator is {\it nothing else but} the {\it formal adjoint} of the {\it Riemann operator}, namely the (linearized) Riemann tensor with $n^2(n^2-1)/2=6$ independent components when $n=3$ [27, 28]. \\
. 
 
 However, if  $\Omega$ is a perturbation of the metric $\omega$, the standard implicit summation used in continuum mechanics is, when $n=3$:  \\
 \[   \begin{array}{rcl}
{\sigma}^{ij}{\Omega}_{ij} & = & {\sigma}^{11}{\Omega}_{11} + 2 {\sigma}^{12}{\Omega}_{12} + 2 {\sigma}^{13}{\Omega}_{13} + {\sigma}^{22} {\Omega}_{22} + 2{\sigma}^{23}{\Omega}_{23} + {\sigma}^{33}{\Omega}_{33}  \\
   &  =  & {\Omega}_{22}d_{33}{\phi}_{11}+ {\Omega}_{33}d_{22}{\phi}_{11}- 2 {\Omega}_{23}d_{23}{\phi}_{11}+ ... \\
   &    & + {\Omega}_{23}d_{13}{\phi}_{12}+{\Omega}_{13}d_{23}{\phi}_{12}- {\Omega}_{12}d_{33}{\phi}_{12}- {\Omega}_{33}d_{12}{\phi}_{12} + ...
\end{array}  \]
because {\it the stress tensor density $\sigma$ is supposed to be symmetric}. Integrating by parts in order to construct the adjoint operator, we get:  \\
\[ \begin{array}{rcl}
 {\phi}_{11} &  \longrightarrow  &  d_{33}{\Omega}_{22} + d_{22}{\Omega}_{33} - 2 d_{23}{\Omega}_{23} \\
 {\phi}_{12} &  \longrightarrow   &  d_{13}{\Omega}_{23}+d_{23}{\Omega}_{13}-d_{33}{\Omega}_{12} - d_{12}{\Omega}_{33}
 \end{array}  \]
and so on. The identifications $ Beltrami = ad(Riemann), \,\, Lanczos = ad(Bianchi)$ in the diagram:  \\
   \[ \fbox{  $  \begin{array}{rcccccccl}  
   & 3  & \underset 1 {\stackrel{Killing}{\longrightarrow}} & 6  & \underset 2 {\stackrel{Riemann}{\longrightarrow}}& 6  & \underset 1 {\stackrel{Bianchi}{\longrightarrow }}  & 3 & \longrightarrow 0  \\
  0 \longleftarrow  & 3  & \underset 1 {\stackrel{Cauchy}{\longleftarrow}} & 6   & \underset 2 {\stackrel{Beltrami}{\longleftarrow}} 
  & 6 &  \underset 1 {\stackrel{Lanczos}{\longleftarrow}} & 3 &    
\end{array}. $   } \]
prove that the $Cauchy$ operator has nothing to do with the $Bianchi$ operator [27, 28]. \\
When $\omega$ is the Euclidean metric, the link between the two sequences is established by means of the elastic constitutive relations $2 {\sigma}_{ij} = \lambda tr(\Omega) {\omega}_{ij} + 2 \mu {\Omega}_{ij}$ with the Lam\'{e} elastic constants $(\lambda, \mu) $ but mechanicians are usually setting $ {\Omega}_{ij} = 2 {\epsilon}_{ij} $. Using the standard Helmholtz decomposition ${\vec{\xi}} = {\vec{\nabla} }\varphi + {\vec{\nabla}} \wedge {\vec{\psi}} $ and substituting in the dynamical equation $ d_i {\sigma}^{ij} =  \rho d^2/dt^2 {\xi}^j $ where $\rho$ is the mass per unit volume, we get the longitudinal and transverse wave equations, namely $ \Delta \varphi - \frac{\rho}{\lambda + 2 \mu} \frac{d^2}{dt^2} \varphi = 0 $ and 
$ \Delta {\vec{\psi}} - \frac{\rho}{\mu} \frac{d^2}{dt^2} {\vec{\psi}} = 0 $, responsible for earthquakes !. \\

Taking into account the factor $2$ involved by multiplying the second, third and fifth row by $2$, we get the new $6\times 6$ operator matrix with rank $3$ which is clearly self-adjoint:  \\
\[  \fbox{  $  \left(  \begin{array}{cccccc}
 0 & 0 & 0 & d_{33} & - 2d_{23} & d_{22} \\
 0 & - 2d_{33} & 2d_{23} & 0 & 2d_{13} & - 2d_{12}  \\
 0 & 2d_{23} & - 2d_{22} & - 2d_{13} & 2d_{12} & 0 \\
 d_{33}& 0 & - 2 d_{13} & 0 & 0 & d_{11}  \\
 - 2d_{23} & 2d_{13} & 2d_{12}& 0 & - 2d_{11} & 0 \\
 d_{22} & - 2 d_{12} & 0 & d_{11}& 0 & 0 
 \end{array} \right)  $  }   \]  \\

{\it Surprisingly}, the Maxwell parametrization is obtained by keeping ${\phi}_{11}=A, {\phi}_{22}=B, {\phi}_{33}=C$ while setting ${\phi}_{12}={\phi}_{23}={\phi}_{31}=0$ in order to obtain the system:\\
\[   \left\{  \begin{array}{rl}
{\sigma}^{11} \equiv& d_{33}B + d_{22}C=0  \\
{\sigma}^{22}\equiv & d_{33}A+ d_{11}C =0 \\
- {\sigma}^{23}\equiv & d_{23}A=0  \\
{\sigma}^{33}\equiv & d_{22}A+ d_{11}B=0  \\
- {\sigma}^{13}\equiv & d_{13}B=0 \\
- {\sigma}^{12}\equiv & d_{12}C=0
\end{array} \right.\]
{\it This system may not be involutive} and no CC can be found "{\it a priori} " because the coordinate system must be changed. Effecting the linear change of coordinates 
${\bar{x}}^1 = x^1, {\bar{x}}^2 = x^2, {\bar{x}}^3 = x^3 + x^2 + x^1 $ and taking out the bar for simplicity, we obtain the involutive system as a {\it  Pommaret basis}:  \\
\[   \left\{  \begin{array}{l}
d_{33}C+ d_{13}C+ d_{23}C+ d_{12}C=0  \\
d_{33}B+ d_{13}B=0  \\
d_{33}A+ d_{23}A=0  \\
d_{23}C +d_{22}C - d_{13}C - d_{13}B - d_{12}C =0  \\
d_{23}A - d_{22}C + d_{13}B + 2 d_{12}C - d_{11}C=0  \\
d_{22}A + d_{22}C - 2 d_{12}C + d_{11}C + d_{11}B=0
\end{array} \right. 
\fbox{ $ \begin{array}{ccc}
1 & 2 & 3   \\
1 & 2 & 3  \\
1 & 2 &  3  \\
1 & 2 &  \bullet  \\
1 &  2 & \bullet  \\
1 &  2 & \bullet
\end{array} $ } \]
and the $3$ CC obtained just amount to the desired $3$ stress equations when coming back to the original system of coordinates. {\it We have thus a minimum parametrization}. Again, {\it if there is a geometrical background, this change of local coordinates is hiding it totally}. Moreover, we notice that the stress functions kept in the procedure are just the ones on which ${\partial}_{33}$ is acting. The reason for such an apparently technical choice is related to very general deep arguments in the theory of differential modules that the extension modules do not depend on the differential sequence used for defining them [20].  \

Finally, setting ${\phi}_{13} = {\phi}_{23} = {\phi}_{33} = 0$, we may even provide the new minimum parametrization:  \\
\[  \fbox{  $  \left\{  \begin{array}{rll}
{\sigma}^{11} \equiv & d_{33}{\phi}_{22} = 0  \\
 - {\sigma}^{12}\equiv &  d_{33}{\phi}_{12} =0  \\
 {\sigma}^{22}\equiv & d_{33}{\phi}_{11} = 0  \\
 {\sigma}^{13}\equiv & d_{23}{\phi}_{12} - d_{13}{\phi}_{22} =0 \\
 - {\sigma}^{23}\equiv &   d_{23}{\phi}_{11} -d_{13} {\phi}_{12} = 0 \\
{\sigma}^{33}\equiv & d_{22}{\phi}_{11}+ d_{11}{\phi}_{22}-2 d_{12}{\phi}_{12} = 0
\end{array} \right.  
\fbox{  $  \begin{array}{ccc}
 1 & 2 & 3  \\
 1 & 2 & 3  \\  
 1 & 2 & 3  \\
 1 & 2 & \bullet  \\
 1 & 2 & \bullet  \\
 1 & 2 & \bullet
 \end{array}  $   }  $    }    \]  \\

\noindent
{\bf PROPOSITION  8.1}: The $Cauchy$ operator can be parametrized by the formal adjoint of the $Ricci$ operator ($4$ terms) and the $Einstein$ operator ($6$ terms) is thus useless. The gravitational waves equations are thus nothing else than the formal adjoint of the linearized $Ricci$ operator.  \\

{\it Proof}:  The {\it Einstein } operator $\Omega \rightarrow E$ is defined by setting $E_{ij}=R_{ij}-\frac{1}{2}{\omega}_{ij}tr(R)$ that we shall 
 write $Einstein= C \circ Ricci$ where $C:S_2T^*\rightarrow S_2T^*$ is a symmetric matrix only depending on $\omega$, which is invertible whenever $n\geq 3$. {\it Surprisingly}, we may also introduce the {\it same} linear transformation $ C:\Omega \rightarrow \bar{\Omega}=\Omega - \frac{1}{2}\omega \, tr(\Omega)$ and the unknown composite operator  ${\cal{X}}: \bar{\Omega}\rightarrow \Omega \rightarrow E$ in such a way that $Einstein= {\cal{X}}\circ C$ where ${\cal{X}}$ is defined by (See [38], 5.1.5 p 134):  \\
 \[  2E_{ij} =   {\omega}^{rs}d_{rs}{\bar{\Omega}}_{ij} - {\omega}^{rs}d_{ri}{\bar{\Omega}}_{sj}-{\omega}^{rs}d_{sj}{\bar{\Omega}}_{ri}+{\omega}_{ij}{\omega}^{ru}{\omega}^{sv}d_{rs}{\bar{\Omega}}_{uv}  \]
Now, introducing the test functions ${\lambda}^{ij}$, we get:  \\
\[   {\lambda}^{ij} E_{ij}={\lambda}^{ij}(R_{ij} -\frac{1}{2}{\omega}_{ij} tr(R)) = ({\lambda}^{ij}-\frac{1}{2} {\lambda}^{rs}{\omega}_{rs}{\omega}^{ij})R_{ij}= {\bar{\lambda}}^{ij}R_{ij}  \]
Integrating by parts while setting as usual $\Box = {\omega}^{rs}d_{rs}$, we obtain:  \\
\[   (\Box {\bar{\lambda}}^{rs}+ {\omega}^{rs}d_{ij}{\bar{\lambda}}^{ij}-{\omega}^{sj}d_{ij}{\bar{\lambda}}^{ri}- {\omega}^{ri}d_{ij}{\bar{\lambda}}^{sj}){\Omega}_{rs}  =  {\sigma}^{rs}{\Omega}_{rs}\]
Moreover, suppressing the "bar " for simplicity, we have:   \\
\[   d_r{\sigma}^{rs}={\omega}^{ij}d_{rij} {\lambda}^{rs}+{\omega}^{rs}d_{rij}{\lambda}^{ij}-
{\omega}^{sj}d_{rij}{\lambda}^{ri} - {\omega}^{ri} d_{rij}{\lambda}^{sj}=0  \]
As $Einstein$ is a self-adjoint operator (contrary to the Ricci operator), we have the identities:   \\
\[  ad(Einstein)=ad(C) \circ ad({\cal{X}}) \,  \Rightarrow  \,  Einstein = C \circ ad({\cal{X}}) \, \Rightarrow  \, ad({\cal{X}})=Ricci \, \Rightarrow \,{\cal{X}}=ad(Ricci)  \]
Indeed, $ad(C) = C$ because $C$ is a symmetric matrix and we know that $ad(Einstein)=Einstein$. Accordingly, the operator $ad(Ricci)$ parametrizes  the $Cauchy$ equations, {\it without any reference} to the $Einstein$ operator that cannot be obtained by any diagram chasing.  The three terms after the {\it Dalembert} operator disappear if we add the {\it differential constraints} $d_i{\lambda}^{ri}=0$. When $n=4$, we finally obtain the adjoint sequences:  \\
 \[  \fbox{  $  \begin{array}{rccccl}
  &4 & \stackrel{Killing}{\longrightarrow } & 10 & \stackrel{Ricci}{\longrightarrow}& 10  \\
   &  &  &  &  &    \\
0 \leftarrow & 4 & \stackrel{Cauchy}{\longleftarrow} & 10 &\stackrel{ad(Ricci)}{\longleftarrow} & 10 
                   \end{array} $  }   \]  
{\it without any reference} to the $Bianchi$ operator and the induced $div$ operator but the upper sequence is not exact because the CC of the {\it Killing} operator are generated by the {\it Riemann} operator, {\it not} by the {\it Ricci} operator as we saw.   \\                                                                                                                                                                                                                                                                                                                                                                                                                                                                                                                        
\hspace*{12cm}  $ \Box  $  \\

\noindent
{\bf REMARK  8.2}: In the opinion of the author of this paper who is not an historian of sciences but a specialist of mathematical physics interested by the analogy existing between {\it electromagnetism} (EM), {\it elasticity} (EL) and {\it gravitation} (GR) by using the conformal group of space-time, it is difficult to imagine that Einstein could not have been aware of the works of Maxwell and Beltrami on the foundations of EL and tensor calculus. Indeed, not only they were quite famous when he started his research work but it must also be noticed that the phenomenological law of field-matter couplings (piezzoelectricity, photoelasticity) had been discovered by ... Maxwell !. \\  \\

In order to extend the classical control concepts of {\it poles} and {\it zeros} to arbitrary $n$, we need a few more definitions.  \\

\noindent
{\bf DEFINITION 8.3}: The symbol $g_q$ of a given system $R_q \subset J_q(E)$ is $g_q = R_q \cap S_q T^* \otimes E \subset J_q(E)$. With any differential module $M$ we shall associate the {\it graded module} $G=gr(M)$ over the polynomial ring $gr(D)\simeq K[\chi]$ by setting $G={\oplus}^{\infty}_{q=0} G_q$ with $G_q=M_q/M_{q+1}$ and we may also set  $g_q=G_q^* = hom_K(G_q, K)$ where the {\it symbol} $g_{q+1}$ is defined by the dual short exact sequences with respect to $K$: \\
\[ 0\longrightarrow M_q\longrightarrow M_{q+1} \longrightarrow G_{q+1} \longrightarrow 0  \hspace{4mm}  \Longleftrightarrow \hspace{4mm}  0 \longrightarrow g_{q+1} \longrightarrow R_{q+1}\stackrel{{\pi}^{q+1}_q}{ \longrightarrow} R_q \longrightarrow 0  \]

\noindent
{\bf PROPOSITION 8.4}: The Spencer operator $d:R_{q+} \rightarrow T^* \otimes R_q$ restricts to $  \delta : g_{q+1} \rightarrow T^* \otimes g_q$ up  to sign and more generally to the so-called $\delta$-sequence of Spencer, namely [1, 2]:
\[ ...{\wedge}^{s-1} T^ \otimes g_{q  +r +1}   \stackrel{\delta}{ \longrightarrow} {\wedge}^s T^* \otimes g_{q+r} \stackrel{\delta}{\longrightarrow}  {\wedge}^{s+1}T^* \otimes g_{q+r-1} ...   \]
We denote by $B^s_{q+r}(g_q)= im(\delta) \subseteq Z^s_{q+r}(g_q) = ker(\delta), H^s_{q+r}(g_q)$ the purely algebraic coboundary, cocycle and cohomology bundles at ${\wedge}^s T^* \otimes g_{q+r}$. The  cohomology is vanishing for any $0 \leq s \leq n$ and for any $r \geq 0$ when $R_q$ and thus $g_q$ is involutive. Also, $g_q$ is said to be $s-acyclic$ if 
$H^1 _{q+r}(g_q)=0, ... , H^s_{q+r}(g_q)=0, \forall r \geq 0$.  \\

Let $R_q \subset J_q(E)$ be a first order involutive system of order $q$ with no CC of order $1$. The following diagram allows to compute the number $dim(F_1)$ of CC of order $2$: \\

\noindent
\[   \begin{array}{rcccccccccl}
  && 0 && 0 &&  0 & & 0  &  & \\
&  &\downarrow  &  &\downarrow  &  & \downarrow  &  &  \downarrow &  &       \\
0 & \rightarrow & g_3 & \rightarrow & S_3 T^* \otimes T & \rightarrow &  S_2T^* \otimes F_0 & \rightarrow & F_1 & \rightarrow & 0  \\
    &  & \downarrow  &  & \downarrow  &  &  \downarrow &  & \parallel &  &  \\
0 & \rightarrow &   R_3  &  \rightarrow & J_3(T) &  \rightarrow & j_2(F_0) & \rightarrow & F_1   &  \rightarrow & 0  \\ 
&  & \downarrow  &  &\downarrow   &  & \downarrow   &  & \downarrow  &  &  \\
0 & \rightarrow &  R_2 & \rightarrow & J_2(T) & \rightarrow & J_1(F_0) & \rightarrow &   0   &  &    \\
   &  & \downarrow  &  & \downarrow  &  &\downarrow   &  &   &  &  \\
   &  & 0 &  & 0 &  & 0 &  &  &  &
\end{array}  \]

We shall apply the previous results to the following inclusion of groups:  \\
\[  \fbox{  $   POINCARE \hspace{2mm} GROUP  \subset CONFORMAL\hspace{2mm} GROUP    $  }  \]
that is  $10 < 15$ when $n=4$ and our aim is now to explain why {\it the mathematical structures  of electromagnetism and gravitation only depend on the second order jets}.  \\

With more details, the Killing system $R_2 \subset J_2(T)$ is defined by the infinitesimal Lie equations in {\it Medolaghi form} with the well known {\it Levi-Civita isomorphism} $(\omega,\gamma )\simeq j_1(\omega)$ involving the metric $\omega$ with $det(\omega)\neq 0$ and the corresponding Christoffel symbols 
${\gamma}^k_{ij}= \frac{1}{2} {\omega}^{kr}({\partial}_i {\omega}_{rj} + {\partial}_j {\omega}_{ir} - {\partial}_r {\omega}_{ij}$. replacing the partial derivatives of vectors by jet notation, we have the defining system:  \\
\[ \left\{  \begin{array}{rcl}
{\Omega}_{ij} & \equiv & {\omega}_{rj}{\xi}^r_i + {\omega}_{ir}{\xi}^r_j + {\xi}^r{\partial}_r{\omega}_{ij}=0  \\
{\Gamma}^k_{ij} & \equiv &  {\gamma}^k_{rj}{\xi}^r_i + {\gamma}^k_{ir}{\xi}^r_j - {\gamma}^r_{ij}{\xi}^k_r + 
     {\xi}^r {\partial}_r{\gamma}^k_{ij} = 0
     \end{array} \right.  \]          
We notice that $R_2(\bar{\omega})=R_2(\omega) \Leftrightarrow \bar{\omega}= a \, \omega, a=cst, \bar{\gamma}=\gamma$ and refer the reader to [LAP] for more details about the link between this result and the deformation theory of algebraic structures. We also notice that $R_1$ is formally integrable and thus $R_2$ is involutive if and only if $\omega$ has constant Riemannian curvature along the well known result of L. P. Eisenhart [  ]. The only structure constant $c$ appearing in the corresponding Vessiot structure equations is such that 
$\bar{c}= c/a$ [CST]. The symbol $g_1$ with $dim(g_1)= n(n-1)/2$ is defined by $ {\omega}_{rj}{\xi}^r_i + {\omega}_{ir}{\xi}^r_j = 0 $ and we have $g_2 = 0$. One can find in any GR textbook the fact that the number of components of the Riemann tensor is $ n^2 (n+1)^2 /4 - n^2 (n+1)(n+2)/6 = n^2 (n^2 - 1) / 12$, thus $20 $ when $n=4$. As we have the short exact sequence:  \\
\[  0 \longrightarrow {\wedge}^2 T^* \otimes g_1 \stackrel{\delta}{\longrightarrow} {\wedge}^3 T^* \otimes  T \longrightarrow 0   \]
the fact that we have also $dim(H^2_1(g_1)) = n^2 (n-1)^2/4 - n^2 (n-1)(n-2)/6 = n^2 (n^2-1)/12$ while changing "$+$" to "$-$" is definitively proving that the foundations of Riemannian geometry must be revisited !. The proof can be obtained by applying the $\delta$-map to the top row of the preceding diagram as it will be done for the conformal case.  \\

 The conformal system ${\hat{R}}_2 \subset J_2(T)$ is defined by the following infinitesimal Lie equations:   \\
 \[   \left \{  \begin{array}{rcl}
   {\omega}_{rj}{\xi}^r_i + {\omega}_{ir}{\xi}^r_j + {\xi}^r{\partial}_r{\omega}_{ij}& =  & 2A(x){\omega}_{ij}   \\
  {\xi}^k_{ij}  + {\gamma}^k_{rj}{\xi}^r_i + {\gamma}^k_{ri}{\xi}^r_j - {\gamma}^r_{ij}{\xi}^k_r  + {\xi}^r{\partial}_r{\gamma}^k_{ij} &  =  & 
  {\delta}^k_i A_j(x) + {\delta}^k_jA_i(x) - {\omega}_{ij}{\omega}^{kr}A_r(x)
 \end{array} \right.    \]
 and is involutive if and only if ${\partial}_iA-A_i= 0$ or, equivalently, if $\omega$ has vanishing Weyl tensor. \\
 Introducing the {\it metric density}  ${\hat{\omega}}_{ij} = {\omega}_{ij}/(\mid det(\omega\mid)^{\frac{1}{n}}$ and substituting, we obtain the system:  \\
 \[  {\hat{\Omega}}_{ij}  \equiv  {\hat{\omega}}_{rj}{\xi}^r_i + {\hat{\omega}}_{ir}{\xi}^r_j - \frac{2}{n}{\hat{\omega}}_{ij}{\xi}^r_r + {\xi}^r{\partial}_r{\hat{\omega}}_{ij} = 0  \]

 Contracting the first equations by ${\hat{\omega}}^{ij}$ we notice that ${\xi}^r_r$ is no longer vanishing. It is essential to notice that the symbols ${\hat{g}}_1$ and ${\hat{g}}_2$ only depend on $\omega$ and not on any conformal factor. Hence, we obtain $dim({\hat{g}}_1) = n(n-1)/2 +1$ and ${\hat{g}}_1$ is simply defined by 
 $ {\omega}_{rj}{\xi}^r_i + {\omega}_{ir}{\xi}^r_j - \frac{2}{n} {\omega}_{ij}{\xi}^r_r  = 0$ with now $dim({\hat{g}}_2) = n$. We have proved in [26] that ${\hat{g}}_3= 0$ when $n \geq 3$, that ${\hat{g}}_2$ is $2$-acyclic when $n \geq 4$ and $3$-acyclic when $n \geq 5$ (These results have been checked by computer algebra up to $n=5$ [39]). \\

 When $n=4$ and ${\hat{g}}_3=0  \Rightarrow {\hat{g}}_4 = 0 \Rightarrow  {\hat{g}}_5 = 0 $ in the conformal case, we have the commutative diagram with exact vertical long $\delta$-sequences {\it but the left one}:

\[  \begin{array}{rcccccccccl}
  & &  0 & & 0 & & 0 &  &  &  & \\
  & & \downarrow & & \downarrow & & \downarrow & & & & \\
0 & \rightarrow & {\hat{g}}_3 & \rightarrow &  S_3T^*\otimes T & \rightarrow & S_2T^*\otimes {\hat{F}}_0& \rightarrow & {\hat{F}}_1 & \rightarrow & 0  \\
  & & \hspace{2mm}\downarrow  \delta  & & \hspace{2mm}\downarrow \delta & &\hspace{2mm} \downarrow \delta & & & & \\
0 & \rightarrow& T^*\otimes {\hat{g}}_2&\rightarrow &T^*\otimes S_2T^*\otimes T & \rightarrow &T^*\otimes T^*\otimes {\hat{F}}_0 &\rightarrow & 0 &&  \\
  & &\hspace{2mm} \downarrow \delta &  &\hspace{2mm} \downarrow \delta & &\hspace{2mm}\downarrow \delta &  &  & &  \\
0 & \rightarrow & {\wedge}^2T^*\otimes {\hat{g}}_1 & \rightarrow & {\wedge}^2T^*\otimes T^*\otimes T & \rightarrow & {\wedge}^2T^*\otimes {\hat{F}}_0 & \rightarrow & 0 &&  \\
 &  &\hspace{2mm}\downarrow \delta  &  & \hspace{2mm} \downarrow \delta  &  & \downarrow  & &  & & \\
0 & \rightarrow & {\wedge}^3T^*\otimes T & =  & {\wedge}^3T^*\otimes T  &\rightarrow   & 0  &  &  &  & \\
 &   &  \downarrow  &  &  \downarrow  &  &  &  &  &  &\\
  &  &  0  &   & 0  & &  &  &  &&
\end{array}  \]

\[  \begin{array}{rcccccccccl}
  & &   & & 0 & & 0 &  &  & &  \\
  & & & & \downarrow & & \downarrow & & &  & \\
   & & 0 & \rightarrow &  80 & \rightarrow & 90 & \rightarrow & 10 & \rightarrow &  0  \\
  &  & \downarrow  & & \hspace{2mm}\downarrow \delta & &\hspace{2mm} \downarrow \delta & & & &  \\
0 & \rightarrow& 16 &\rightarrow & 160& \rightarrow & 144 &\rightarrow & 0 &   \\
 &  &\hspace{2mm} \downarrow \delta &  &\hspace{2mm} \downarrow \delta & &\hspace{2mm}\downarrow \delta &  &  &  &  \\
0 & \rightarrow & 42 & \rightarrow & 96 & \rightarrow & 54 & \rightarrow & 0 & &  \\
  & &\hspace{2mm}\downarrow \delta  &  & \hspace{2mm} \downarrow \delta  &  & \downarrow  & & & & \\
0 & \rightarrow & 16 & =  & 16 &\rightarrow   & 0  &  &  &  &  \\
  &  &  \downarrow  &  &  \downarrow  &  &  &  &  &  & \\
  &  &  0  &   & 0  & &  &  &  & &
\end{array}  \]

A diagonal snake chase proves that ${\hat{F}}_1 \simeq H^2 ({\hat{g}}_1)$. However, we obtain at once $dim(B^2({\hat{g}}_1))=  16$ but, in order to prove that the number of components of the Weyl tensor is $42-32= 10$ or, equivalently, to prove that $dim(Z^2 ({\hat{g}}_1))=42 - 16 = 26$, we have to prove that the last map $\delta$ in the left Weyl $\delta$-sequence is surjective, a result that it is almost impossible to prove in local coordinates. Let us prove it by means of circular diagram chasing in the preceding commutative diagram as follows. Lift any $a \in {\wedge}^3 T^* \otimes T$ to $b \in {\wedge}^2 T^* \otimes T^* \otimes T$ because the vertical $\delta$-sequence for $S_3 T^*$ is exact. Project it by the symbol map ${\sigma}_1({\hat{\Phi}})$ to $c \in {\wedge}^2 T^* \otimes {\hat{F}}_0$. Then, lift $c$ to $d \in T^* \otimes T \otimes {\hat{F}}_0$ that we may lift {\it backwards horizontally} to 
$e \in T^* \otimes S_2 T^* \otimes T$ to which we may apply $\delta$ to obtain $f \in {\wedge}^2 T^* \otimes T^* \otimes T$. By commutativity, both $f $ {\it and} $b$ map to $c$ and the difference $f - b$ maps thus to zero. Finally, we may find $g \in {\wedge}^2 T^* \otimes {\hat{g}}_1 $ such that $ b = g + \delta (e)$ and we obtain thus $a = \delta (g) + {\delta}^2 (e) = \delta (g)$, proving therefore the desired surjectivity. Going one step further, we let the reader discover the following result, found in 2016 but still not acknowledged today !: \\
WHEN $n=4$, THE WEYL TENSOR WITH $10$ COMPONENTS HAS ONLY $9$ GENERATING BIANCHI-LIKE CC OF ORDER $2$; \\
In order to prove that both classical and conformal differential geometry must be entirely revisited, let us prove that the analogue of the Weyl tensor is made by a third order operator when $n=3$ which is also neither known nor nacknowledged today. As before, we shall proceed by diagram chasing as the local computation can only be done by using computer algebra and does not provide any geometric insight (See [39] for the details).  \\

\[   \begin{array}{rcccccccccl}
  & & 0  &  & 0  & &  0  &           \\
  & & \downarrow  &  &  \downarrow & & \downarrow &    \\
0  & \rightarrow & {\hat{g}}_4 & \rightarrow &S_4T^*\otimes T& \rightarrow &S_3T^ *\otimes {\hat{F}}_0 & \rightarrow & {\hat{F}}_1 & \rightarrow  &  0 \\
& & \downarrow  &  &  \downarrow & & \downarrow & &    \\
0 & \rightarrow & T^*\otimes {\hat{g}}_3 & \rightarrow &T^*\otimes S_3T^*\otimes T& \rightarrow &T^*\otimes S_2T^ *\otimes {\hat{F}}_0 &  \rightarrow &0 & &   \\
& & \downarrow  &  &  \downarrow & & \downarrow & & &    \\
0 & \rightarrow &{\wedge}^2 T^*\otimes {\hat{g}}_2 & \rightarrow &{\wedge}^2T^* \otimes S_2T^* \otimes T& \rightarrow & {\wedge}^2T^* \otimes T^* \otimes {\hat{F}}_0 &\rightarrow & 0 & &    \\
& & \downarrow  &  &  \downarrow & & \downarrow & & &   \\
0 & \rightarrow &{\wedge}^3 T^*\otimes {\hat{g}}_1 & \rightarrow &{\wedge}^3T^*\otimes T^*\otimes T& \rightarrow &{\wedge}^3T^*\otimes {\hat{F}}_0 &\rightarrow &0& \\
&   & \downarrow  &  &  \downarrow & & \downarrow & & &    \\
0 & \rightarrow &{\wedge}^4 T^*\otimes T & =  &{\wedge}^4T^*\otimes T& \rightarrow &  0 &  &  &  \\
 &&  \downarrow  &  &  \downarrow & & & & &    \\
 &&  0  &  & 0  & &    &  &   &        
\end{array}  \]

\[   \begin{array}{rcccccccccl}
  & &   &  & 0  & &  0  &           \\
  & &   &  &\downarrow   & & \downarrow &    \\
  &  & 0 & \rightarrow &  45  & \rightarrow &  50 & \rightarrow & 5 & \rightarrow  &  0 \\
  & &   &  &  \downarrow & & \downarrow & &    \\
  &  & 0 & \rightarrow &  90 & \rightarrow &  90 &  \rightarrow &0 & &   \\
& & \downarrow  &  &  \downarrow & & \downarrow & & &    \\
0 & \rightarrow & 9 & \rightarrow & 54 & \rightarrow & 45 &\rightarrow & 0 & &    \\
& & \downarrow  &  &  \downarrow & & \downarrow &  &  \\
0 & \rightarrow & 4  & \rightarrow & 9 & \rightarrow &  5   &\rightarrow &0& \\
&   & \downarrow  &  &  \downarrow &  &  \downarrow & & &     \\
 &&  0  &  & 0  & & 0   &  &   &        
\end{array}  \]
We have indeed: \,$3\, translations\, + \, 3 \, rotations \, +  \, 1 \, dilatation \  + 3 \, elations \, = 10 $ \, parameters.\\
We obtain from a chase ${\hat{F}}_1= H^2({\hat{g}}_1)$ and the totally unexpected formally exact sequences on the jet level are thus, showing in particular that second order CC do not exist:  \\
\[  0 \rightarrow {\hat{R}}_3 \rightarrow J_3(T)  \rightarrow J_2({\hat{F}}_0) \rightarrow 0 \,\, \Rightarrow \,\, 
0 \rightarrow 10 \rightarrow 60 \rightarrow 50 \rightarrow 0  \] 
\[     0  \rightarrow {\hat{R}}_4 \rightarrow J_4(T)  \rightarrow J_3({\hat{F}}_0) \rightarrow {\hat{F}}_1 \rightarrow 0 \,\,\, \Rightarrow \,\, 
0 \rightarrow 10 \rightarrow 105 \rightarrow 100 \rightarrow 5 \rightarrow 0 \]
\[     0  \rightarrow {\hat{R}}_5 \rightarrow J_5(T)  \rightarrow J_4({\hat{F}}_0) \rightarrow J_1({\hat{F}}_1) \rightarrow {\hat{F}}_2  \rightarrow 0 \,\,\, \Rightarrow \,\, 
0 \rightarrow 10 \rightarrow 168 \rightarrow 175 \rightarrow 20 \rightarrow   3    \rightarrow 0 \]
We obtain the minimum differential sequence, {\it which is nervertheless  not a Janet sequence}:

\[   0 \longrightarrow {\hat{\Theta}} \longrightarrow T \underset 1{\stackrel{{\hat{\cal{D}}}}{\longrightarrow}} {\hat{F}}_0 \underset 3{\stackrel{{\hat{\cal{D}}}_1}{\longrightarrow}} {\hat{F}}_1 \underset 1{\stackrel{{\cal{D}}_2}{\longrightarrow} } {\hat{F}}_2  \longrightarrow 0 \,\,  \, \Rightarrow \,\, \,
0 \longrightarrow {\hat{\Theta}} \longrightarrow 3 \underset 1{\stackrel{{\hat{\cal{D}}}}{\longrightarrow}} 5 \underset 3{\longrightarrow} 5 \underset 1{\longrightarrow} 3 \rightarrow 0   \]
with $\hat{\cal{D}}$ the conformal Killing operator and vanishing Euler-Poincar\'{e} characteristic $3 - 5 + 5 - 3 = 0$. We have proved very recently ({\it to appear soon}) that the $5 \times 5$ operator  $ {\hat{\cal{D}}}_1$ is self-adjoint.   \\

As a byproduct, we end this paper with the following {\it fundamental diagram} $II$ first presented in $1983$ [18] but still not yet acknowledged 
as it only depends on the Spencer $\delta$-map, explaining both the splitting vertical sequence on the right  existing in Riemannian geometry and the vector bundle isomorphism 
${\it Ricci} \simeq S_2T^*$. Indeed, the diagonal chase providing it could not be even imagined by using classical methods because its involves Spencer $\delta$-cohomology with the standard notations $B=im(\delta), Z= ker(\delta), H=Z/B$ for {\it coboundary, cocycle, cohomology} at ${\wedge}^sT^* \otimes g_{q+r}$ when $g_{q+r}$ is the $r$-prolongation of a symbol $g_q$. It is important to notice that all the bundles appearing in this diagram only depend on the metric $\omega$ but {\it not} on any conformal factor. \\
 
\[  \fbox{  $   \begin{array}{rcccccccccl}
&  &  &  &  &  &  &  &  &  & \\
  &  &  &  &  &  &   &  &  0  & & \\
  &  &  &  &  &  &   &  & \downarrow &  & \\
  &  &  &  &  &  &  0  &  &  Ricci & &  \\
  &  &  &  &  &  &  \downarrow &  & \downarrow &  &  \\
  &  &  &  &  0 & \longrightarrow & Z^2_1(g_1) & \longrightarrow & Riemann & \longrightarrow & 0  \\
  &  &  &  &   \downarrow &  & \downarrow &  & \downarrow &  &  \\
  &  & 0 & \longrightarrow & T^* \otimes {\hat{g}}_2 & \stackrel{\delta}{\longrightarrow} & Z^2_1({\hat{g}}_1) & \longrightarrow & Weyl  & \longrightarrow & 0  \\
  &  &  &  &  \downarrow &  &  \downarrow &  &  \downarrow &  &  \\
 0 &  \longrightarrow & S_2T^* &  \stackrel{\delta}{\longrightarrow} &  T^* \otimes T^* & \stackrel{\delta}{\longrightarrow} & {\wedge}^2 T^* &  \longrightarrow & 0 &  &  \\      
  &  &  &  & \downarrow &  &  \downarrow &  &  &  &  \\
  &  &  &  &  0 &  &  0  &  &  &  &\\
  &  &  &  &  &  &  &  &  &  & 
  \end{array}  $  }  \]   \\

When $n=4$, we have explained in recent books [8, 39] and papers [28, 32], that the horizontal lower sequence is splitting because $a_{i,j}=\frac{1}{2}(a_{i,j} + a_{j,i}) + \frac{1}{2} (a_{i,j} - a_{j,i})$ and provides an isomorphism $T^* \otimes {\hat{g}}_2 \simeq T^* \otimes T^* \simeq S_2T^* \oplus {\wedge}^2 T^*$ which can be locally described by $ (R_{ij}, F_{ij})$ in which $(R_{ij})$ is the GR part and $(F_{ij})$ the EM part as a unification of gravitation and electromagnetism. Accordingly, the bundle splitting $Riemann \simeq Ricci \oplus Weyl$ is thus only depending on the second order jets of conformal transformations, {\it contrary to the philosophy of GR today}. We finally notice that such a result is contradicting the mathematical foundations of classical gauge theory relating EM to $U(1)$ while allowing to understand the confusion done by E. Cartan and followers between "{\it curvature alone}" ($F_1$) and "{\it curvature + torsion}" ($C_2 = {\wedge}^2 T^* \otimes R_2$).  \\

\noindent
{\bf 9) CONCLUSION}:\\

    We have already proved in our book "{\it Lie Pseudogroups and Mechanics}" (1988) that the group foundation of elasticity pioneered by E. and F. Cosserat (1909) was just described by adding to the {\it non-linear Spencer sequence} (1972) a convenient variational calculus. As a crucial conclusion and similarly to classical elasticity, even if the initial background is non-linear, the resulting stress/couple-stress equations are linear. Accordingly, in order to parametrize these equations, one just needs to refer to {\it infinitesimal Lie equations} and the three corresponding {\it canonical linear differential sequences} that can be constructed, namely the {\it gauge sequence}, the {\it Janet sequence} and the {\it Spencer sequence}, though only the last one is useful.\\
    
    The main result of this paper, coming from {\it unavoidable arguments of homological algebra}, is that the Cosserat couple-stress equations are just described by the formal adjoint of the first Spencer operator while the formal adjoint of the second Spencer operator just describes a possible parametrization. Accordingly, the parametrization of the Cosserat couple-stress equations is {\it first order} while the parametrization of the classical Cauchy stress equations is {\it second order} as it comes from dualizing second order CC described by the Riemann operator. When $n=4$, we have also proved that the use of differential double duality applied to the Spencer sequence for the conformal Lie pseudogroup allows to unify the Cosserat, Maxwell and Einstein equations, hence elasticity, electromagnetism and gravitation, along the dream of H. Weyl and that such a result only depends on the second ordrer jets called "elations" by E. Cartan in $1920$. However, we have proved that the mathematical theory of gravitational waves is not coherent with double differential duality, a {\it fact} explaining why Einstein himself has not been very confident in their existence all along his life.  \\
    
   Nevertheless, the most surprising result is that the same homological arguments, namely the systematic use of the adjoint operators, is also providing the need to revisit the mathematical foundations of control theory. Indeed, when a differential field $K$ having $n$ commuting derivations is given together with two finitely generated differential extensions $L$ and $M$ of $K$, an important problem in differential algebra is to exhibit a common differential extension $N$ in order to define the new differential extensions $L\cap M$ and the smallest differential field $(L,M)\subset N$ containing both $L$ and $M$. Such a result allows to generalize the use of complex numbers in classical algebra. Similarly, having now two finitely generated differential modules $L$ and $M$ over the non-commutative ring ring $D=K[d_1,... ,d_n]=K[d]$ of differential operators with coefficients in $K$, we may look for a differential module $N$ containing both $L$ and $M$ in order to define $L\cap M$ and $L+M$. This is {\it exactly} the situation met in linear or non-linear OD or PD control theory by selecting the inputs and the outputs among the system variables. However, in many recent books and papers, we have shown that controllability was a {\it built-in} property of a control system, not depending on the choice of inputs and outputs among the system variables. Another purpose of this paper is thus to revisit the mathematical foundations of control theory by showing the specific importance of the two previous problems and the part plaid by $N$ in both cases for the parametrization of the control system. The essential tool will be the study of {\it differential correspondences}, a modern name for what was called {\it B\"{a}cklund problem} during the last century, namely the study of elimination theory for groups of variables among systems of linear or nonlinear OD or PD equations. The difficulty is to revisit {\it differential homological algebra} by using non-commutative localization. Finally, when $M$ is a $D$-module, this paper is using for the first time the fact that the system $R=hom_K(M,K)$ is a $D$-module for the Spencer operator acting on sections, avoiding thus behaviors, trajectories and signal spaces. We finally hope that the many explicit applications presented will be used in the future as test examples for computer algebra.  \\  \\  \\  \\

\noindent
{\bf REFERENCES}  \\

\noindent
[1] Pommaret, J.-F.: Systems of Partial Differential Equations and Lie Pseudogroups, Gordon and Breach, New York (1978); Russian translation: MIR, Moscow,(1983).\\
\noindent
[2] Pommaret, J.-F.: Partial Differential Equations and Group Theory, Kluwer (1994).\\
http://dx.doi.org/10.1007/978-94-017-2539-2    \\
\noindent
[3] Pommaret, J.-F.: Dualit\'{e} Diff\'{e}rentielle et Applications, Comptes Rendus Acad\'{e}mie des Sciences Paris, S\'{e}rie I, 320 (1995) 1225-1230.  \\
\noindent
[4] Kashiwara, M.: Algebraic Study of Systems of Partial Differential Equations, M\'{e}moires de la Soci\'{e}t\'{e} Math\'{e}matique de France, 63 (1995) (Transl. from Japanese of his 1970 Master Thesis).  \\
\noindent
[5] Pommaret, J.-F.: Differential Galois Theory, Gordon and Breach, New York (1983).\\
\noindent
[6] Pommaret, J.-F.: Partial Differential Control Theory, Kluwer, Dordrecht (2001) (Zbl 1079.93001).   \\
\noindent
[7] Zerz, E.: Topics in Multidimensional Linear Systems Theory, Lecture Notes in Control and Information Sciences, Springer, LNCIS 256 (2000). \\
\noindent
[8] Pommaret, J.-F.: New Mathematical Methods for Physics, Mathematical Physics Books, Nova Science Publishers, New York (2018) 150 pp.  
https://doi.org/10.4236/jmp.2022.134036   \\
\noindent
[9] Pommaret, J.-F., Quadrat, A.: Localization and Parametrization of Linear Multidimensional Control Systems, Systems \& Control Letters, 37 (1999) 247-260.  \\
\noindent 
[10] Pommaret, J.-F.: Algebraic Analysis of Control Systems Defined by Partial Differential Equations, in "Advanced Topics in Control Systems Theory", Springer, Lecture Notes in Control and Information Sciences 311 (2005) Chapter 5, pp. 155-223.\\
\noindent
[11] Spencer, D.C.: Overdetermined Systems of Partial Differential Equations, Bull. Am. Math. Soc., 75 (1965) 1-114.\\
\noindent
[12] Goldschmidt, H.: Prolongations of Linear Partial Differential Equations: I Inhomogeneous equations, Ann. Scient. Ec. Norm. Sup., 4 (1968) 617-625.  https://doi.org/10.24033/asens.1173\\
\noindent
[13] Palamodov, V. P.: Linear Differential Operators with Constant Coefficients, Springer, 1970.  \\
\noindent
[14] Oberst, U.: Multidimensional Constant Linear Systems, Acta Applicandae Mathematica, 20 (1990) 1-175. https://doi.org/10.1007/BF00046908  \\
\noindent
[15]  Pommaret, J.-F.: Parametrization of Cosserat Equations, Acta Mechanica, 215 (2010) 43-55.\\
http://dx.doi.org/10.1007/s00707-010-0292-y  \\
\noindent
[16] Cosserat, E., Cosserat, F.: Th\'{e}orie des Corps D\'{e}formables, Hermann, Paris (1909).\\
\noindent
[17] Pommaret, J.-F.: Fran\c{c}ois Cosserat and the Secret of the Mathematical Theory of Elasticity, Annales des Ponts et Chauss\'ees, 82 (1997) 59-66 
(Translation by D.H. Delphenich).  \\
\noindent
[18] Pommaret, J.-F.: Lie Pseudogroups and Mechanics, Gordon and Breach, New York (1988).\\
\noindent
[19] Northcott, D.G.: An Introduction to Homological Algebra, Cambridge university Press (1966).  \\
\noindent
[20] Rotman, J.J.: An Introduction to Homological Algebra, Pure and Applied Mathematics, Academic Press (1979).  \\
\noindent
[21] Janet, M.: Sur les Syst\`{e}mes aux D\'{e}riv\'{e}es Partielles, Journal de Math., 8 (1920) 65-151. \\
\noindent 
\noindent
[22] Macaulay, F.S.: The Algebraic Theory of Modular Systems, Cambridge Tract 19, Cambridge University Press, London, 1916 (Reprinted by Stechert-Hafner Service Agency, New York, 1964).  \\
\noindent
[23] Pommaret, J.-F.: Spencer Operator and Applications: From Continuum Mechanics to Mathematical Physics, in "Continuum Mechanics-Progress in Fundamentals and Engineering Applications", Dr. Yong Gan (Ed.), ISBN: 978-953-51-0447--6, InTech (2012) Available from: \\
http://dx.doi.org/10.5772/35607   \\
\noindent
[24] Pommaret, J.-F.: The Mathematical Foundations of Gauge Theory Revisited, Journal of Modern Physics, 5 (2014) 157-170.   
https://doi.org/10.4236/jmp.2014.55026   \\
\noindent
[25] Weyl, H.: Space, Time, Matter, (1918) (Dover, 1952).  \\
\noindent
[26] Pommaret, J.-F.: From Thermodynamics to Gauge Theory: Th Virial Theorem Revisited, in " Gauge Theories and Differential Geometry", Lance Bailey (Ed.), ISBN: 978-1-63483-546-6, Nova Science publishers (2016), New York, Chapter 1, 1-44.  \\
\noindent
[27] Pommaret, J.-F.: Why Gravitational Waves Cannot Exist, Journal of Modern Physics, 8 (2017) 2122-2158.  https://arxiv.org/abs/1708.06575 . 
https://doi.org/104236/jmp.2017.813130    \\
\noindent
[28] Pommaret, J.-F.: Gravitational waves and Lanczos Potentials, Journal of Modern Physics, 14 (2023) 1177-1202. 
https://doi.org/10.4236/jmp.2023.148065  \\  
\noindent
[29] Pommaret, J.-F.: Killing Operator for the Kerr Metric, Journal of Modern Physics, 14 (2023) 31-59. https://doi.org/10.4236/jmp.2023.141003     \\
\noindent
[30] Pommaret, J.-F.: Relative Parametrization of Linear Multidimensional Systems, Multidim. Syst. Sign. Process., 26 (2015) 405-437.  
https://doi.org/10.1007/s11045-013-0265-0   \\
\noindent
[31] Pommaret, J.-F.: How Many Structure Constants do Exist in Riemannian Geometry, Mathematics in Computer Science,     
https://doi.org/10.1007/s11786-022-00546-3   \\
\noindent
[32] Pommaret, J.-F.: The Mathematical Foundations of General Relativity Revisited, Journal of Modern Physics, 4 (2013) 223-239.
 https://dx.doi.org/10.4236/jmp.2013.48A022   \\
 \noindent
 [33] Choquet-Bruhat, Y.: Introduction to General Relativity, Black Holes and Cosmology, Oxford University Press (2015).  \\
\noindent
 [34] Airy, G.B.:  On the Strains in the Interior of Beams, Phil. Trans. Roy. Soc. London, 153 (1863) 49-80.  \\  
\noindent
[35] Beltrami, E.: Osservazioni sulla Nota Precedente, Atti della Accademia Nazionale dei Lincei Rend., 1, 5 (1892) 141-142; Collected Works, t IV . \\
\noindent
[36] Maxwell, J.C.: On Reciprocal Figures, Frames and Diagrams of Forces, Trans. Roy. Soc. Ediinburgh, 26 (1870) 1-40.  \\
\noindent
[37] Morera, G.: Soluzione generale della equazioni indefinite di equilibrio di un corpo continuo, Atti della Academia Nazionale dei Lincei Rend., 1, 5 (1892) 137-141 + 233 - 234.        \\
\noindent
[38] Foster, J., Nightingale, J.D.: A Short Course in General Relativity, Longman (1979).  \\
\noindent
[39] Pommaret, J.-F.: Deformation Theory of Algebraic and Geometric Structures, Lambert Academic Publisher (LAP), Saarbrucken, Germany (2016). 
http://arxiv.org/abs/1207.1964  \\

\end{document}